\documentclass[10pt,a4paper]{article}
\usepackage[margin=2.4cm]{geometry}
\usepackage[T1]{fontenc}
\usepackage{lmodern}
\usepackage{amsmath,amssymb,amsthm,booktabs,graphicx,xcolor,caption,longtable,array}
\usepackage[round]{natbib}
\usepackage[hidelinks]{hyperref}
\newtheorem{proposition}{Proposition}
\theoremstyle{definition}\newtheorem{definition}{Definition}
\newcommand{\indep}{\perp\!\!\!\perp}
\newcommand{\Gs}{\mathcal G^{*}}
\newcommand{\Sh}{\hat\Sigma}
\newcommand{\Pp}{\mathbb P}
\newcommand{\pa}{\mathrm{pa}}
\newcommand{\Ft}{\mathcal F_t}
\newcommand{\f}[1]{\texttt{\detokenize{#1}}}
\newcolumntype{L}[1]{>{\raggedright\arraybackslash}p{#1}}
\newcommand{\BIC}{\mathrm{IC}}

\title{Graphical identification of structural vector autoregressions\\ through lag exclusions: consistency and reliability}
\author{Marco Reale$^{1}$, Anna Lin$^{2}$, Granville Tunnicliffe Wilson$^{3}$,\\ Chris Price$^{1}$ and Blair Robertson$^{1}$\\[4pt]
{\small $^{1}$School of Mathematics and Statistics, University of Canterbury, New Zealand}\\
{\small $^{2}$Statistics New Zealand \quad $^{3}$Department of Mathematics and Statistics, Lancaster University, UK}}
\date{}
\begin{document}
\maketitle

\begin{abstract}
\noindent A directed acyclic graph over the current variables supplies zero restrictions that identify a
structural vector autoregression (SVAR). Learned from data, the identified objects are the arrows compelled in
its Markov equivalence class under the time ordering, and lagged variables do much of the identifying: lag
exclusions orient contemporaneous arrows that the reduced-form innovations leave free. We study this identification and its reliability. Equivalence is defined on the window graph with
the lag distribution unrestricted. For fixed dimension and lag order, faithfulness conditional on the other
lagged variables suffices for the penalised-likelihood minimiser and a dynamic programme over orderings, as
implemented, to recover the class consistently from one stationary series. With growing dimension, a per-parent extended
Bayesian information criterion (BIC) selects each parent set consistently along a true ordering, whereas the
BIC can overfit; in simulations it removed 40--90\,\% of the BIC's errors for 15--50 variables. It is
fragile: with 28 of 42 lag cells non-zero, 148 of 150 arrows were identifiable but only
49\,\% of reported arrows were correct at $n=500$, because missed lags create spurious v-structures. In an
application we show which lag exclusions each fitted arrow rests on. The R package \texttt{gmsvar} implements
the method.
\end{abstract}




%=====================================================================
\section{Introduction}

A vector autoregression (VAR) describes how a vector of time series evolves from its own past. Its innovations
are generally correlated, so it says nothing about how the variables influence each other within a period. A
structural VAR (SVAR) separates these contemporaneous effects from the lagged ones, but the structure is not
identified by the data alone: many structural forms imply the same reduced form. The conventional remedy, a
recursive ordering of the variables, requires prior knowledge that is rarely available, and impulse responses
can change substantially with the order \citep{demiralp2003}.

Graphical modelling offers a data-driven alternative: a directed acyclic graph (DAG) over the current
variables supplies the identifying zero restrictions \citep{swanson1997,demiralp2003,moneta2008}. Earlier work
of this group estimated a conditional independence graph of the current and lagged variables by testing
partial correlations and then enumerated the DAGs compatible with it \citep{reale2001,oxley2009}, a procedure
that is difficult to automate.

When the DAG is learned from data, the lagged variables do much of the identifying. A lagged parent of a
current variable that is not a parent of its contemporaneous parent forms a collider that orients the
arrow between them; a lagged parent of the parent that is not a parent of the child orients the arrow by
preventing a collider. Either way, lag exclusions orient contemporaneous arrows which the reduced-form
innovations leave free. In the designs of this paper, fitting the
window of current and lagged variables identifies more than three times as many arrows as an analysis of the
VAR residuals. Every learner that fits lagged and contemporaneous links together relies on this mechanism, but it
rests on the lag exclusions being right, and its theory and reliability have received little systematic
study. This paper
asks what lag exclusions identify in an SVAR, under which conditions a penalised-likelihood estimator recovers
it from a single dependent series, and how reliable the resulting orientations are.

The estimator uses established ingredients. Search over orderings, with the best parent sets chosen for each
ordering, is that of \citet{teyssier2005}; our local search for larger systems is of this kind, its moves
relocating one variable to its best position rather than swapping adjacent variables, as in the best order
score search (BOSS) of \citet{andrews2023}, and the dynamic programme over subsets is that of
\citet{silander2006}. Our penalty is a per-parent modification of the extended BIC of \citet{chen2008}. Residual-based graphical identification of SVARs
\citep{swanson1997,demiralp2003,moneta2008} is established. Closest to our problem, \citet{ahelegbey2016}
identify SVARs by Bayesian graphical models, sampling the contemporaneous and the lagged structure by MCMC;
the R package BiDAG \citep{suter2023} provides order-based MCMC for dynamic Bayesian networks with
within-slice and between-slice edges; DYNOTEARS \citep{pamfil2020} estimates both sets of coefficients by
penalised least squares under a smooth acyclicity constraint; PCMCI$^{+}$ \citep{runge2020} learns
contemporaneous and lagged links by conditional independence tests; and \citet{bang2023} and
\citet{larsen2025} characterise, and learn by a score-based search, the equivalence class restricted by
temporal tiers. Two learners for time series are closer still. TS-BOSS \citep{castillo2026}, developed concurrently with
and independently of this work, also searches
over permutations of the current variables with the lagged variables placed first, choosing parent sets by
grow-shrink trees and refining the result by a backward equivalence search; its consistency argument treats
the window observations as independent and identically distributed (its Assumption~5). The SVAR-GFCI
procedure of \citet{malinsky2018}, designed for latent confounding, starts from a greedy equivalence search
constrained by the time order and by homologous (repeated) edges. For reduced-form VARs, \citet{lingtu2025}
screen with the extended BIC and select lags consistently with a modified BIC.

A search over current-variable permutations with lagged candidate parents is therefore not new: TS-BOSS
has one, and with the same score the two searches give similar graphs (Section~\ref{sec:learners}). The search
is our vehicle, not our contribution. It uses the conditional Gaussian likelihood of the current variables
given the lags, with free innovation variances and no restriction on the joint distribution of the lagged
variables; all lagged variables are eligible parents of every current variable; for up to ten variables the
search is exhaustive over orderings; and the output is the equivalence class under the time ordering. Our
contributions are the following.
\begin{enumerate}\setlength{\itemsep}{0pt}
\item \emph{The identification target} (Sections~\ref{sec:setting} and~\ref{sec:theory}): equivalence of SVAR
DAGs on the augmented window graph, with the distribution of the lagged variables unrestricted; what a
compelled arrow and an invariant parent set identify; the rule, a consequence of mixed-time colliders and Meek's first rule, that a contemporaneous link can be
reversible only if its ends have the same lagged parents, which we develop and evaluate; and the observation
that faithfulness conditional on the other lagged variables suffices for recovering the class, whereas
faithfulness over all conditioning sets can fail without any cancellation.
\item \emph{Consistency for one dependent series} (Section~\ref{sec:theory}): of the penalised-likelihood
minimiser (Proposition~\ref{prop:min}) and of the dynamic programme with backward elimination as implemented
(Proposition~\ref{prop:be}), for a fixed number of variables and lag order, using only the convergence of the
sample covariance of the window; existing guarantees for TS-BOSS assume a consistent score with i.i.d.\
windows.
\item \emph{The penalty} (Sections~\ref{sec:theory} and~\ref{sec:large}): as the number of lagged candidates
grows, the BIC overfits, through null lagged variables that each lower it, while a per-parent extended BIC
selects each parent set consistently along a true ordering (Proposition~\ref{prop:grow}); in simulation the
extended BIC removed 40 to 90\,\% of the errors from fifteen to fifty variables.
\item \emph{Reliability of orientation through lags} (Sections~\ref{sec:lags} and~\ref{sec:macro}): as the lag
structure becomes denser or weaker, missed lagged links create spurious v-structures, and the share of
reported arrows that are correct falls well before the true directions stop being identifiable; in an
application we find, for each fitted arrow, the smallest set of lagged links whose addition would make it
reversible, and show that the instability of the orientations under resampling comes from the selection of
the lags.
\item \emph{Comparisons} (Section~\ref{sec:results}): with published learners, with TS-BOSS and arc-wise
search under an identical score, and with search on the VAR residuals.
\end{enumerate}

%=====================================================================
\section{Setting}\label{sec:setting}

\subsection{Structural VAR and window}\label{sec:window}
Let $X_t=(X_{1,t},\dots,X_{k,t})'$ be $k$ time series, measured from their population means, and let $p$ be
the lag order. We assume the structural form
\begin{equation}\label{eq:svar}
A_0X_t=A_1X_{t-1}+\dots+A_pX_{t-p}+\varepsilon_t,\qquad \varepsilon_t\overset{\text{iid}}{\sim}N(0,D),\quad
D=\mathrm{diag}(\sigma_1^2,\dots,\sigma_k^2),
\end{equation}
with $A_\ell\in\mathbb R^{k\times k}$ for $\ell=0,\dots,p$ and all $\sigma_i^2>0$, and take the stationary
causal solution of \eqref{eq:svar}; the fitted regressions contain a free intercept (Section~\ref{sec:method}),
so a non-zero mean is allowed in the data. $A_0$ has unit diagonal and is, after a permutation of the
variables, lower triangular, so that the contemporaneous relations are acyclic. The reduced form is
$X_t=\sum_{\ell}\Phi_\ell X_{t-\ell}+e_t$, with $\Phi_\ell=A_0^{-1}A_\ell$ and reduced-form innovation
$e_t=A_0^{-1}\varepsilon_t$, $\operatorname{Var}(e_t)=A_0^{-1}DA_0^{-\prime}$. It is stable when all zeros of
$\det(I_k-\sum_{\ell=1}^p\Phi_\ell z^\ell)$ lie outside the unit circle, equivalently when the companion
matrix of $\Phi_1,\dots,\Phi_p$ has spectral radius below one; we report the latter. Many structural forms
share a reduced form. A recursive ordering resolves the choice by the $k(k-1)/2$ zeros of an
upper-triangular part of $A_0$ after permuting the variables, and a specified sparse DAG can impose further
zeros; when the DAG is learned, the target is the class of Definition~1 below. Equation \eqref{eq:svar} is
read as a DAG on the window vector
\[
Z_t=(X_t',X_{t-1}',\dots,X_{t-p}')'\in\mathbb R^m,\qquad m=k(p+1),\qquad Z_{j+\ell k,t}=X_{j,t-\ell}
\quad(j\le k,\ 0\le\ell\le p),
\]
whose first $k$ coordinates are the current variables and the others, $L=\{k+1,\dots,m\}$, the lagged ones;
$Z_{P,t}$ is the subvector indexed by $P\subseteq\{1,\dots,m\}$, and we drop $t$ when convenient. A
non-zero $(A_0)_{ij}$, $i\ne j$, is an edge $X_{j,t}\to X_{i,t}$ and a non-zero $(A_\ell)_{ij}$ an edge
$X_{j,t-\ell}\to X_{i,t}$. The structural equation of $Z_i$, $i\le k$, is
$Z_{i,t}=\sum_{j\in\pa(i)}\beta_{ij}Z_{j,t}+\varepsilon_{i,t}$, with $\beta_{ij}=-(A_0)_{ij}$ for a
contemporaneous parent $j\le k$ and $\beta_{i,j+\ell k}=(A_\ell)_{ij}$ for $\ell\ge1$; the coefficients of
Figure~\ref{fig:macro} follow this convention. Only current variables have parents that are modelled:
$\pa(i)\subseteq\{1,\dots,m\}\setminus\{i\}$ for $i\le k$. Two modelling choices are made throughout: \emph{all
lagged variables are eligible parents} of every current variable, and the model places \emph{no restriction on
the joint distribution of the lagged variables}. We call such a structure an SVAR DAG. From a sample
$X_1,\dots,X_n$ there are $N=n-p$ window observations.

\subsection{Equivalence under the time ordering}\label{sec:equiv}
The skeleton of a DAG is the undirected graph obtained by removing its arrowheads, and a v-structure is a
triple $a\to c\leftarrow b$ with $a$ and $b$ non-adjacent. Two DAGs are Markov equivalent if and only if they
have the same skeleton and the same v-structures \citep{verma1990,andersson1997}. Here the orientation of every lag-to-current edge is known from the arrow of
time. The joint distribution of the lagged variables is unrestricted (Section~\ref{sec:window}), and the
conditional likelihood of Section~\ref{sec:method} does not model it. We encode both choices in the \emph{augmented graph} of
an SVAR DAG, obtained by adding all edges among lagged vertices (oriented, say, by index); two lagged parents
of a variable are then never an unshielded pair. Leaving the joint distribution of the lagged variables unrestricted is a modelling
device: it forgoes any information in conditional independencies among them.

\begin{definition}
Two SVAR DAGs on the same window are equivalent if their augmented graphs have the same skeleton and the same
set of v-structures.
\end{definition}

An edge is \emph{compelled} if it has the same orientation in every equivalent DAG. Only contemporaneous
edges can fail to be compelled. Starting from the skeleton and the v-structures, we add the orientations given
by the time order and apply the rules R1--R3 of \citet{meek1995}; rule R4 is not needed with tiered background
knowledge \citep{bang2023}. Equivalence classes, not individual DAGs, are the target of identification: SVAR DAGs in the
same class, lagged and contemporaneous parts together, attain the same maximised conditional likelihood and
have the same number of edges.

\emph{Example: one lag exclusion identifies an arrow.} Let $W=X_{t-1}$ and
\[
X_t=\varepsilon_{X,t},\qquad Y_t=bX_t+cW+\varepsilon_{Y,t},\qquad bc\ne0,
\]
with independent Gaussian innovations. Given the past, the innovations of $X_t$ and $Y_t$ are $\varepsilon_{X,t}$
and $b\varepsilon_{X,t}+\varepsilon_{Y,t}$, and their covariance is fitted equally well by either recursive
ordering, so the VAR residuals cannot orient the link. In the window graph, however, $X_t\to Y_t\leftarrow W$
is an unshielded collider. Reversing the contemporaneous arrow while keeping the skeleton gives the chain
$W\to Y_t\to X_t$, which implies $W\perp X_t\mid Y_t$, whereas the model has $W\perp X_t$ and
$W\not\perp X_t\mid Y_t$. The arrow $X_t\to Y_t$ is therefore compelled, and it is compelled only because
$W\to X_t$ is absent: if that edge is included in the graph the triangle is complete and both orientations
fit, the enlarged model representing the original distribution with a zero coefficient on the added edge.
Allowing the edge as a candidate in the search does not by itself shield the collider. A missed
lagged link can thus turn a reversible link into an apparently compelled arrow.

What is identified follows from the class. The structural equation of $X_{i,t}$ is the regression of $X_{i,t}$ on its parents, so its
coefficients, its innovation variance and its shock $\varepsilon_{i,t}$ are the same in every DAG of the class
when its parent set is the same in every DAG, that is, when every contemporaneous link at $X_{i,t}$ is compelled. The responses to
$\varepsilon_{i,t}$ at all horizons are then identified too, since the $i$th column of $A_0^{-1}$ is
$\operatorname{Cov}(e_t,\varepsilon_{i,t})/\operatorname{Var}(\varepsilon_{i,t})$ with $e_t$ the reduced-form
innovation. A compelled arrow into a variable that also has a reversible link has an identified direction
but, in general, not an identified coefficient, and the responses to that variable's shock depend on the
member of the class.

The example generalises. A lagged parent of $X_{i,t}$ that is not a parent of a contemporaneous parent
$X_{j,t}$ forms a v-structure that compels $X_{j,t}\to X_{i,t}$, and a lagged parent of $X_{j,t}$ that is not
a parent of $X_{i,t}$ orients $X_{j,t}\to X_{i,t}$ by Meek's first rule. A contemporaneous link can therefore
be reversible only if its two ends have the same lagged parents. Both mechanisms, mixed-time colliders and
Meek's first rule, are used by \citet[Appendix~B]{castillo2026}; the rule is a consequence that we develop and
evaluate (Sections~\ref{sec:lags} and~\ref{sec:macro}). Across the 20 designs with seven variables each of Section~\ref{sec:design}, the true window graphs compel
all 150 contemporaneous arrows, whereas the contemporaneous graphs alone, which are all that an analysis of
the VAR residuals can use, compel 42. This is a comparison of population classes; Sections~\ref{sec:results}
and~\ref{sec:lags} give the finite-sample recovery. The gain rests on the lag exclusions being correct, a
condition examined in Section~\ref{sec:lags}.

%=====================================================================
\section{Method}\label{sec:method}

\subsection{Score}
For an SVAR DAG $\mathcal G$ with parent sets $\pa(i)$, $i\le k$, the Gaussian likelihood conditional on the
first $p$ observations factors over dates and, within each date, over the structural equations: because
$\det A_0=1$ the change of variables from $\varepsilon_t$ to $X_t$ has unit Jacobian, and the
log-likelihood is, up to a constant, $-\frac12\sum_{i\le k}\{N\log\sigma_i^2+\mathrm{RSS}_i/\sigma_i^2\}$.
Profiling out the $\sigma_i^2$ leaves $N\sum_i\log\hat\sigma_i^2\{\pa(i)\}$ up to a constant, which is
decomposable, and so is the penalised score
\begin{equation}\label{eq:score}
\mathrm{IC}(\mathcal G)=\sum_{i=1}^k s_i\{\pa(i)\},\qquad s_i(P)=N\log\hat\sigma_i^2(P)+|P|\lambda_N.
\end{equation}
Here every regression contains a free intercept, and $\hat\sigma_i^2(P)$ is the maximum-likelihood residual
variance, the residual sum of squares divided by $N$, of the regression of $Z_i$ on $Z_P$, computed from the
centred sample covariance matrix $\Sh$ of the window (divisor $N$). Its population counterpart
$\sigma_i^2(P)$ is the residual variance of the population regression of $Z_i$ on $Z_P$; the innovation
variance is $\sigma_i^2=\sigma_i^2\{\pa^*(i)\}$ for the true parent set. With $\lambda_N=\log N$ the score is
the BIC. Stability is not imposed in the search: enforcing it would couple the equations and destroy
decomposability, so it is checked after the fit (Section~\ref{sec:macro}). When each variable has many
candidate parents the BIC overfits at moderate $n$ (Section~\ref{sec:large}). The extended BIC of
\citet{chen2008} adds $2\gamma\log\binom{q}{|P|}$ for a parent set chosen among $q$ candidates. With fixed
candidate sets this term is also additive over nodes, but $q$ changes with the ordering. We replace it by
$2\gamma|P|\log(m-1)$, with $m-1$ the number of possible parents of a current variable, a per-parent
modification that gives the penalty
\begin{equation}\label{eq:ebic}
\lambda_N=\log N+2\gamma\log(m-1),\qquad\gamma\ge0,
\end{equation}
a constant cost per parent across nodes and orderings. We call \eqref{eq:score} with \eqref{eq:ebic} the
extended BIC and use $\gamma=0.5$, the default of the software; $\gamma=0$ is the BIC. Equivalent structures
have the same score.

\subsection{Search over all orderings}
Every acyclic contemporaneous structure admits a topological ordering of the $k$ current variables. For
$U\subseteq\{1,\dots,k\}$ let $\mathrm{best}(U)$ be the minimum of $\sum_{i\in U}s_i\{\pa(i)\}$ over
acyclic structures on $U$ with $\pa(i)\subseteq[(U\setminus\{i\})\cup L]\cap\mathcal N(i)$ for each $i\in U$.
Then
\begin{equation}\label{eq:dp}
\mathrm{best}(\emptyset)=0,\qquad
\mathrm{best}(U)=\min_{i\in U}\bigl[\mathrm{best}(U\setminus\{i\})+s\{i,C(i,U\setminus\{i\})\}\bigr],\qquad
s(i,C)=\min_{P\subseteq C}s_i(P),
\end{equation}
where, for a set $S$ of current predecessors of $i$, $C(i,S)=(S\cup L)\cap\mathcal N(i)$ is the set of
candidate parents: the current variables already placed and all lagged variables, intersected with an
optional set $\mathcal N(i)$ of admissible parents of $i$ (Section~\ref{sec:restrict}). The recursion is the dynamic programme of \citet{silander2006}, with the arrow of time fixing the
position of the lagged variables. In our implementation (Algorithm~1) the minimisation over $P\subseteq C$ is
replaced by backward elimination: starting from $P=C$, the candidate whose deletion gives the lowest score is
deleted as long as this lowers the score. The starting fit with $q=|C|$ candidates and an intercept needs
$N>q+1$ and a non-singular candidate covariance; in the unrestricted search $q\le m-1$, so it needs
$N=n-p>k(p+1)$. When $q\ge N-1$ the software uses forward selection from the empty set instead, and when the
candidate covariance is numerically singular it solves each regression directly, a singular set scoring
$+\infty$. With $q$ candidates, one Cholesky factorisation and rank-one
downdates make a node problem $O(q^3)$, so the programme costs $O(k2^km^3)$: exponential in the number of
current variables only. A fit takes about 0.1\,s for $k=7$, $p=3$ and 1\,s for $k=10$, $p=2$ on one core, and
the software uses this search up to $k=10$. We call it the \emph{full search}. It is exhaustive over
orderings, but backward elimination can return a parent set whose score exceeds $s(i,C)$, so it need not
return the minimiser of \eqref{eq:score}; Proposition~\ref{prop:be} shows that it is nevertheless consistent.

\begin{center}
\fbox{\begin{minipage}{0.92\textwidth}\small
\textbf{Algorithm 1.} Full search: dynamic programme over orderings with backward elimination.\\[2pt]
\textbf{Input:} window covariance $\Sh$ of dimension $m=k(p+1)$, $N$, $k$, lagged indices
$L=\{k+1,\dots,m\}$, penalty $\lambda_N$, admissible parents $\mathcal N(i)$ (all variables by default).\\
1. $\mathrm{best}(\emptyset)\leftarrow0$.\\
2. For $U\subseteq\{1,\dots,k\}$, $U\neq\emptyset$, in order of increasing size:\\
\hspace*{1.5em}for $i\in U$: $C\leftarrow\{(U\setminus\{i\})\cup L\}\cap\mathcal N(i)$;\quad
$P_i\leftarrow$ backward elimination on $s_i$ from $C$ (memoised on $(i,C)$);\\
\hspace*{1.5em}$\mathrm{best}(U)\leftarrow\min_{i\in U}\{\mathrm{best}(U\setminus\{i\})+s_i(P_i)\}$, recording
the minimising $i$ and its $P_i$.\\
3. Return the parent sets obtained by tracing the minimisers back from $U=\{1,\dots,k\}$.
\end{minipage}}
\end{center}

\subsection{Local search for larger systems}
For $k>10$ the parent sets are found as above and the ordering is improved by moving one variable at a time
to its best position, as in BOSS \citep{andrews2023}, node problems being memoised on the set of
predecessors. The greedy start places next, among the variables not yet placed, the one whose score improves
least when the variables already placed become candidates. Each pass visits the variables in random order and
moves each to the position that gives the lowest total score, if this lowers the score by more than
$10^{-9}$ (among equal positions the first is taken); a run ends after a pass without a move, or after 50
passes. Three restarts from random permutations follow the greedy start, and the ordering with the lowest
score is returned, a later run replacing an earlier one only if its score is strictly lower. Against the full search it returned the same parent sets in 40 of 40 fits in Model~1, 118 of
120 in R7, 59 of 60 on random designs with $k=8$ to 10 and 12 of 12 on series from the $k=15$ design
(Supplement). A fit takes about 2\,s at $k=15$ and several minutes at $k=50$. This search is a heuristic; the
results of Section~\ref{sec:theory} do not cover it.

\subsection{Restricting the candidates}\label{sec:restrict}
The sets $\mathcal N(i)$ allow prior knowledge, or a first-stage screen of the window such as a graph of
significant partial correlations, to restrict the search. They restrict only the DAG that is fitted: the
equivalence class, and with it the compelled status of each arrow, is computed from the fitted DAG under the
time ordering alone, so a restriction never compels an arrow. A data-dependent screen is not prior knowledge,
and a directional restriction would call for a class that includes it, which we do not compute. A screen is
not needed for identification, and the search cannot restore a true parent that a screen has dropped. The
default uses no screen, and all results below are for the unrestricted search unless stated.

%=====================================================================
\section{Theory}\label{sec:theory}

\subsection{A fixed number of variables}
Let $\Gs$ be the true SVAR DAG and $\pa^*(i)$ its parent sets.

\begin{proposition}[Penalised-likelihood minimiser]\label{prop:min}
Let $k$ and $p$ be fixed and let $Z_t$ be the window vector of a stable Gaussian SVAR with DAG $\Gs$ and $D$
positive definite. Assume faithfulness given the past: for $i\le k$, every $j\in\{1,\dots,m\}\setminus\{i\}$
and every $S$ with $L\setminus\{j\}\subseteq S\subseteq\{1,\dots,m\}\setminus\{i,j\}$, $Z_i\indep Z_j\mid Z_S$
if and only if $i$ and $j$ are d-separated by $S$ in the augmented graph of $\Gs$. Let $\lambda_N\to\infty$ and
$\lambda_N/N\to0$. Let $\mathcal N_N(i)$ be possibly data-dependent sets of admissible parents with
$\Pp\{\pa^*(i)\subseteq\mathcal N_N(i)\text{ for all }i\}\to1$, let $\mathcal S_N$ be the set of SVAR DAGs
with $\pa(i)\subseteq\mathcal N_N(i)$ for every $i$, and let $\hat{\mathcal G}$ minimise \eqref{eq:score}
over $\mathcal S_N$. Then $\Pp\{\hat{\mathcal G}\text{ is equivalent to }\Gs\}\to1$.
\end{proposition}

The conditioning sets contain every lagged variable other than $j$. Faithfulness over all conditioning
sets would fail without any cancellation of coefficients: for two independent AR(1) series $X$ and $Y$,
$X_t\indep Y_{t-1}$, although the path $Y_{t-1}-X_{t-1}\to X_t$ is open in the augmented graph; given
$X_{t-1}$ it is blocked. Given all other lagged variables, every path through a lagged vertex other than an endpoint is blocked.
The condition still involves the lagged variables: when $j$ is lagged it concerns the lag-to-current
adjacencies, conditioning on the other lags. The proof (Appendix) uses it only
through a weaker consequence: every SVAR DAG whose model contains the true window covariance has at least
$|\Gs|$ edges, with equality only in the class of $\Gs$, a sparsest-representation condition in the sense of
\citet{raskutti2018}; that condition can replace faithfulness in the proposition. The proof uses only the
convergence in probability of the sample covariance of the window and its $\sqrt N$-consistency
\citep{lutkepohl2005}, so it covers dependent observations, and it holds for the BIC and for the extended BIC
with fixed $\gamma$; it fixes the lag order at its true value. It concerns the exact minimiser. The next result removes that
restriction for the search that the software uses for up to ten variables.

\begin{proposition}[Full search with backward elimination]\label{prop:be}
Under the assumptions of Proposition~\ref{prop:min}, let $\hat{\mathcal G}_{\mathrm{BE}}$ be the structure
returned by Algorithm~1, in which for every node $i$ and candidate set $C$ the parent set is obtained by
backward elimination: starting from $P=C$, a candidate whose deletion lowers $s_i$ is deleted, and the
procedure stops when no single deletion lowers $s_i$. Then
$\Pp\{\hat{\mathcal G}_{\mathrm{BE}}\text{ is equivalent to }\Gs\}\to1$.
\end{proposition}

Backward elimination need not return the best subset for every node and candidate set, and the proposition
does not say that it does: on series from our designs it did so in 94 to 100\,\% of the node problems in
Model~1 and in 63 to 95\,\% in R7, with little effect on mean skeleton error, although in R7 at $n=200$ the
fitted graph was the same as with exact parent sets in only 18\,\% of the fits (Supplement). The proposition says that
along a true topological ordering it returns the true parent sets with probability tending to one. The
dynamic programme then attains a score no larger than that of $\Gs$, and by Proposition~\ref{prop:min} only
structures equivalent to $\Gs$ can do so.

\subsection{A growing number of variables}
Propositions~\ref{prop:min} and~\ref{prop:be} keep $k$ fixed, whereas the simulations show the BIC overfitting
as $k$ grows and the extended BIC removing most of the excess. The next result explains this for the
parent-selection step. Fix the contemporaneous ordering at a topological ordering of $\Gs$ and, for each
current variable $i$, let $C_i$ be its candidates (the variables before it in the ordering and all lagged
variables), $S_i\subseteq C_i$ its true parents with the coefficients $\beta_{ij}$ of Section~\ref{sec:window},
and, for a fixed integer cap $K$ on the number of parents, $\hat S_i$ the minimiser of $s_i(P)$ over
$P\subseteq C_i$ with $|P|\le K$. Now $k=k_N$ and $m=m_N=k_N(p+1)$ may grow with $N$, with $p$
fixed. Assume:
\begin{itemize}\setlength{\itemsep}{0pt}
\item[(G1)] $X$ is a stable Gaussian SVAR of dimension $k_N$ whose window covariance has all eigenvalues in a
fixed interval $[\underline\lambda,\bar\lambda]\subset(0,\infty)$ and the largest eigenvalue of whose
spectral density matrix is bounded above uniformly over frequencies and $N$ (with $p$ fixed, this bounds the
spectrum of the window up to a factor $p+1$);
\item[(G2)] $m_N\to\infty$ and $m_N-1\le cN^{\kappa}$ for some $c>0$ and $0<\kappa<1$, and $\gamma\ge0$;
\item[(G3)] $|S_i|\le\bar s\le K$ for fixed $\bar s$ and $K$, and $\min_{i,\,j\in S_i}|\beta_{ij}|\ge b_N$ with
$Nb_N^2/\log N\to\infty$.
\end{itemize}

\begin{proposition}[Extended BIC and BIC with many candidates]\label{prop:grow}
Under (G1)--(G3), with the ordering fixed as above:
\begin{itemize}\setlength{\itemsep}{0pt}
\item[(a)] if $\gamma>1-1/(2\kappa)$, then $\max_{i\le k_N}\Pp(\hat S_i\ne S_i)\to0$;
\item[(b)] if $\gamma>2-1/(2\kappa)$, then $\Pp(\hat S_i=S_i\text{ for all }i\le k_N)\to1$;
\item[(c)] suppose that for some $i$ there is a set $R_N$ of lagged candidates that are not parents of $i$,
with $|R_N|\ge N^{\kappa'}$, whose residuals given $Z_{S_i}$ have pairwise correlations at most
$\bar\rho\in[0,1)$. For $j\in R_N$ let $t_j$ be the signed coefficient of $Z_j$ divided by its usual standard
error in the regression of $Z_i$ on an intercept, $Z_{S_i}$ and $Z_j$, and suppose that
$\max_{j\in R_N}t_j$ admits a Gaussian approximation (assumption (GA), stated exactly in Section~S1). If
$2\kappa'(1-\bar\rho)>1+2\gamma\kappa$, then with probability tending to one no minimiser of (2) over all SVAR
DAGs is equivalent to $\Gs$, and the same holds over the SVAR DAGs compatible with any fixed topological
ordering of $\Gs$; for the size-capped estimator $\hat S_i$ it holds when $|S_i|<K$.
\end{itemize}
\end{proposition}

The threshold in (a) is that of \citet{chen2008} for a single regression, carried over to a stable VAR; (b)
adds a union over nodes. \citet{lingtu2025} prove screening and selection consistency of extended- and
modified-BIC procedures for the equations of a reduced-form VAR; the extension here is to the selection of
structural parent sets, contemporaneous and lagged, along a given true ordering of the current variables. Along a true topological ordering, each candidate regressor is known before the node's innovation is
revealed, so the sums of regressor--innovation products that drive the scores are martingales with
conditionally Gaussian increments, and their tails are as sharp as with independent observations. In the regime of (c) with $|R_N|$ of order $N^{\kappa}$, $\bar\rho=0$ and (GA), and in the example below,
the condition in (a) is close to necessary: the largest spurious $\chi^2_1$ statistic among $N^{\kappa}$ null
candidates is about $2\kappa\log N$, against a penalty of $(1+2\gamma\kappa)\log N$. With fewer or correlated
null candidates, (c) gives a sufficient condition for overfitting that does not match the threshold of (a). Under the conditions of (c) with $\bar\rho=0$, the BIC
therefore overfits when the number of such candidates grows like $N^{\kappa'}$ with $\kappa'>1/2$, and by (a)
the extended BIC with $\gamma=0.5$ selects each parent set consistently whenever the number of candidates
grows like $N^{\kappa}$ with $\kappa<1$. These are statements about polynomial rates: a number of candidates
such as $N/\log N$ is smaller than $N$ but is not covered. Part (b), recovery of all parent sets at once, needs a larger $\gamma$ because $k_N$ grows
with $m_N$ (for $\gamma=0.5$ it requires $\kappa<1/3$), which is consistent with a few spurious links remaining
in the largest designs. The proposition idealises the parent step of the software, which imposes no bound on
$|P|$ and uses backward elimination, and it takes the ordering as given: the search over orderings, whose
number grows like $k!$, needs further identifiability conditions \citep{vandegeer2013,nandy2018} and is not
covered. The proof is in the Supplement.

\emph{Example: BIC overfitting without a Gaussian approximation.} Let $Y_t=\phi Y_{t-1}+\varepsilon_t$ with
$0<|\phi|<1$ and $\varepsilon_t$ Gaussian white noise, and let $W_{1,t},\dots,W_{q,t}$ be Gaussian white-noise
series, independent of one another and of $Y$ (hence of $\varepsilon$), with $q=q_N\asymp N^{\kappa}$ and
$0<\kappa<1$. This is a stable SVAR with $p=1$, $k=q+1$ and no contemporaneous links, and
$m-1\asymp N^\kappa$. Consider adding one lagged candidate $W_{j,t-1}$ to the true regression of $Y_t$ on an
intercept and $Y_{t-1}$, and let $t_j$ be the signed $t$-statistic of its coefficient. Given the whole $Y$ series, the columns $(W_{j,t-1})_t$ are independent isotropic
Gaussian vectors; after projection off the intercept and $Y_{t-1}$ their directions are uniform in the
residual space, so the $q$ statistics $t_j$ are conditionally independent with the exact $t_{N-3}$
distribution, and $\max_jt_j^2/\log N\to2\kappa$ in probability. Adding $W_{j,t-1}$ changes the score by
$\lambda_N-N\log\{1+t_j^2/(N-3)\}=\lambda_N-t_j^2\{1+o_p(1)\}$. With the BIC, $\lambda_N=\log N$, so for
$\kappa>1/2$ some spurious lagged edge improves on the true graph with probability tending to one; with the
extended BIC, $\lambda_N=(1+2\gamma\kappa)\log N\{1+o(1)\}$, and no single addition does once
$\gamma>1-1/(2\kappa)$, the threshold of part (a). A growing collection of individually unremarkable lagged
variables is enough to defeat the BIC.

%=====================================================================
\section{Simulation design}\label{sec:design}

\emph{Model 1} is an SVAR(2) in $k=3$ variables ($m=9$) with the contemporaneous edges $X_1\to X_2$ and
$X_1\to X_3$, three non-zero entries in each lag matrix and innovation variances $(0.09,0.04,0.025)$; its
matrices are given in Section~S2 of the Supplement.
\emph{R7} is a set of 20 random sparse stable SVAR(3) designs in $k=7$ variables ($m=28$): each variable
after the first in a random ordering receives $\min\{r-1,B\}$ contemporaneous parents, drawn at random among its $r-1$ predecessors, where
$B\sim$ Binomial$(3,0.5)$ and $r$ is its position (a draw larger than $r-1$ is capped) (4 to 10 edges, 7.5 on average, with v-structures); the lag-1 matrix has an autoregressive
diagonal in $[0.35,0.6]$ and 12 off-diagonal entries in $\pm[0.15,0.35]$, and the matrices at lags 2 and 3
have 9 entries each in $\pm[0.1,0.25]$; the spectral radius of the companion matrix is 0.83 to 0.89. Many
links are weak. The population partial correlation of a pair given all other window variables is below 0.10
in absolute value for 8.6 of the 44.5 true edges of a design on average; among the pairs with a current
variable, 70 such partial correlations are non-zero per design on average, co-parents included, and 26 of
them are below 0.10. \emph{K15} is one random sparse
SVAR(2) in $k=15$ variables ($m=45$) with 19 contemporaneous edges. \emph{Larger designs} are random
SVAR(2)s with $k=20$ (18 designs), 25 (12), 30 (6) and 50 (12), with one series per design and sample size;
a variable then has between 59 ($k=20$) and 149 ($k=50$) possible parents over all orderings, and $kp$
lagged candidates plus its predecessors under a fixed ordering. Section~\ref{sec:lags} varies the lag structure
of R7. Innovations are Gaussian and series are generated after a burn-in of 500.

The true DAG and its equivalence class are read from $A_0,\dots,A_p$. We report the number of \emph{skeleton
errors}, the missing plus extra edges among the pairs that involve a current variable (contemporaneous pairs
and lag-to-current cells: 21 in Model~1, 168 in R7, 555 in K15), the share of runs in which the estimate is
equivalent to the truth, and, for orientation, the share of the truly compelled arrows that the estimate
compels in the right and in the wrong direction. The true lag order is used unless stated. The Supplement
indexes every study with its designs, seeds, replications, search, penalty and software version; the same
cell can appear with slightly different values in two tables because they use different series.

%=====================================================================
\section{Results}\label{sec:results}

\subsection{Small systems}

Table~\ref{tab:small} gives the full search without a restriction of the candidates and, as a benchmark
that is not available in practice, restricted to the true moral graph: the parents of each common child are
joined, the arrowheads removed, and the candidates $\mathcal N(i)$ of a current variable $i$ are its
neighbours in that graph. Without any restriction the search with the extended BIC recovers the class of
Model~1 in 0.74 of the runs at $n=100$ and 0.94 at $n=500$, close to what the true moral graph would allow,
and clearly more often than with the BIC at small $n$ (0.52 and 0.76 at $n=100$ and 200). In R7 the class is
recovered in at most 0.10 of the runs at $n\le500$ even with the oracle graph, because many true links are
weak, and in 0.44 (BIC), 0.64 (extended BIC) and 0.71 (oracle) at $n=2000$. Of the skeleton errors, which
also count wrong lagged links, most are among the 147 lag-to-current cells: at $n=500$, 8.2 and 8.6 of the 8.6
and 8.8 errors without a restriction and 4.9 of the 5.0 with the oracle, the 21 contemporaneous pairs
contributing 0.4, 0.2 and 0.04. A graph estimated from the data is not a substitute for the oracle:
restricting the search to one costs accuracy in these designs, the more so the more conservative the rule
(Supplement). In R7 the two penalties have about equal mean skeleton errors.

\begin{table}[t]\centering\small
\begin{tabular}{llrrr}
\toprule
 & $n$ & BIC & Extended BIC & Oracle graph, BIC\\
\midrule
\multicolumn{5}{l}{\emph{Model 1: share of runs recovering the true class}}\\
 & 100 & 0.52 & 0.74 & 0.89\\
 & 200 & 0.76 & 0.91 & 0.95\\
 & 500 & 0.84 & 0.94 & 0.95\\
 & 2000 & 0.96 & 0.99 & 0.99\\
\multicolumn{5}{l}{\emph{R7: mean skeleton errors among 168 cells}}\\
 & 200 & 19.4 (1.1) & 19.3 (0.9) & 11.6 (0.2)\\
 & 500 & 8.6 (0.4) & 8.8 (0.2) & 5.0 (0.0)\\
 & 2000 & 1.3 (0.1) & 0.9 (0.0) & 0.6 (0.0)\\
 & 10\,000 & 0.4 (0.1) & 0.1 (0.0) & 0.1 (0.0)\\
\multicolumn{5}{l}{\emph{R7: share of runs recovering the true class}}\\
 & 500 & 0.03 & 0.02 & 0.10\\
 & 2000 & 0.44 & 0.64 & 0.71\\
 & 10\,000 & 0.71 & 0.94 & 0.95\\
\bottomrule
\end{tabular}
\caption{Full search in small systems: unrestricted, with the BIC and the extended BIC ($\gamma=0.5$), and
restricted to the true moral graph (oracle). In parentheses: errors among the 21 contemporaneous pairs.
200 replications for Model~1 and 100 for R7.}\label{tab:small}
\end{table}

\subsection{Comparison with published learners}\label{sec:learners}

\begin{table}[t]\centering\small
\begin{tabular}{lrrrr}
\toprule
 & Model 1 & R7 & R7 & $k=15$\\
 & $n=500$ & $n=500$ & $n=2000$ & $n=1000$\\
\midrule
Our search, extended BIC & 0.2 & 8.5 & 0.8 & 3.4\\
Our search, BIC & 0.3 & 8.8 & 1.0 & 6.3\\
\midrule
PCMCI$^{+}$ (tigramite), $\alpha=0.01$ & 5.9 & 24.6 & 18.5 & 22.4\\
Tiered PC (tpc), $\alpha=0.01$ & 8.4 & 30.7 & 28.8 & 33.8\\
TGES, BIC & 1.3 & 17.9 & 17.8 & 16.1\\
DYNOTEARS, one setting & 5.5 & 14.6 & 11.8 & 14.9\\
DYNOTEARS, best setting per cell & 1.4 & 14.6 & 10.9 & 14.9\\
Hill climbing, BIC (bnlearn) & 0.1 & 14.5 & 12.0 & 16.4\\
Tabu search, BIC (bnlearn) & 0.1 & 14.3 & 11.9 & 13.8\\
\midrule
PC (pcalg), $\alpha=1/\sqrt n$, no time order & 8.0 & 30.1 & 28.1 & 33.4\\
GES, BIC (pcalg), no time order & 3.2 & 30.8 & 34.8 & 49.0\\
\midrule
Series & 20 & 20 & 20 & 10\\
\bottomrule
\end{tabular}
\caption{Published learners on the same series: mean skeleton errors among the scored cells, true lag order.
R7 has one series per design. At $k=15$ tpc, pcalg and bnlearn were run on the first five series, on which
our search made 7.4 (BIC) and 3.8 (extended BIC) errors. DYNOTEARS: penalties 0.01 to 0.1 and weight
thresholds 0 to 0.1; ``one setting'' is penalty 0.1 and threshold 0.05, ``best'' is chosen with knowledge of
the truth.}\label{tab:learners}
\end{table}

Published implementations were run on series from the same designs (Table~\ref{tab:learners}). All made more
skeleton errors than our search in every cell but Model~1, where the arc-wise searches of bnlearn did
slightly better. PC and GES, applied to the window without the time order, are baselines that differ from
ours in the algorithm as well as in ignoring the time order.
TGES \citep{larsen2025}, run with its authors' code and lags as tiers, and DYNOTEARS \citep{pamfil2020},
even with its penalty and threshold chosen with knowledge of the truth, made about two to twenty times as many
errors as our search with the extended BIC. DYNOTEARS also orients the contemporaneous edges much less
reliably when the series are standardised, since it is not invariant to scale (Supplement). Each learner was run in one configuration, or a small grid, on 20
series or fewer per cell, so these results describe performance in our designs and settings, not a ranking
of the method classes.

TS-BOSS \citep{castillo2026}, the closest method, was compared separately with its authors' Python code
(commit d0cf643), which implements the search over orderings of the algorithm but not its backward
equivalence search. Both searches were given the same window matrices and lag orders, and the penalty of
TS-BOSS was set to ours, the BIC or the extended BIC, so that only the search differed. The series were 100
from R7 and 100 from the designs of Table~\ref{tab:lags} with 28 non-zero cross-lag cells (fitted with $p=3$),
at $n=500$ and 2000, and those of the 18 designs with $k=20$ of Table~\ref{tab:large} at $n=300$, 1000 and
3000, where our search is the local one (Supplement). The two searches were close. Mean skeleton errors
differed by at most 1.6; with standard errors computed across designs, the difference exceeded two standard
errors only in R7 at $n=500$, where our search made 0.70 and 1.33 fewer errors with the two penalties
(standard errors 0.34 and 0.55), and TS-BOSS made fewer errors in some cells, for example at $k=20$ and
$n=1000$ with the BIC (12.5 against 13.2); the precision
and recall of the compelled contemporaneous directions differed by at most seven percentage points. Our search
reached the lower score in 3 to 83\,\% of the series of a cell and TS-BOSS in 0 to 44\,\%, the rest tying:
its grow--shrink parent selection sometimes beats our backward elimination even where our dynamic programme
is exhaustive over orderings. A fit took 0.1 against 0.4\,s at $k=7$ and 9 to 12 against 7 to 39\,s at
$k=20$, the implementations being in different languages.

BiDAG \citep{suter2023}, whose dynamic networks are first order, was compared on twenty seven-variable
designs with one lag. With its default score and MCMC settings, learning the transition structure, it was
slightly less accurate than our search with the extended BIC (3.7 and 0.9 errors at $n=500$ and 2000 against
2.3 and 0.5) and about two hundred times slower (Supplement). It is the closest of the learners compared, as
one expects of another order-based search with a consistent score; our search differs in allowing any lag
order, in its speed and in returning the equivalence class.

\subsection{Search and penalty separated}\label{sec:crossed}

\begin{table}[t]\centering\small
\begin{tabular}{llrrrrrrr}
\toprule
 & & & \multicolumn{2}{c}{Search over orderings} & \multicolumn{2}{c}{Arc-wise, equal time} &
\multicolumn{2}{c}{Arc-wise, single run}\\
Design & $n$ & Designs & BIC & ext.\ BIC & BIC & ext.\ BIC & BIC & ext.\ BIC\\
\midrule
R7 & 500 & 20 & 7.4 & 8.0 & 17.3 & 15.8 & 18.6 & 17.8\\
R7 & 2000 & 20 & 1.8 & 1.3 & 10.6 & 8.7 & 12.2 & 10.2\\
$k=20$ & 300 & 18 & 40.4 & 23.5 & 53.6 & 37.7 & 54.8 & 42.5\\
$k=20$ & 1000 & 18 & 13.6 & 6.1 & 33.6 & 25.4 & 40.7 & 31.7\\
$k=20$ & 3000 & 18 & 6.8 & 1.6 & 34.9 & 28.8 & 38.4 & 30.7\\
$k=30$ & 1000 & 6 & 26.8 & 8.0 & 47.7 & 21.0 & 54.7 & 38.5\\
$k=50$ & 1000 & 12 & -- & 11.7 & -- & 14.4 & -- & --\\
\bottomrule
\end{tabular}
\caption{Search type and penalty with one common score and search space: mean skeleton errors, one series per
design. Search over orderings: local search without restarts ($k\le30$; in R7 the full search gives 7.4, 7.7
and 1.8, 1.3) and with three restarts ($k=50$). Arc-wise: hill climbing by addition, deletion or reversal of
one arc from the empty graph (single run), and restarted from random perturbations until the time taken by the
search over orderings is used up (equal time; for $k\le10$ the time of the full search; the last restart is
completed, so the elapsed time can exceed this budget slightly).}\label{tab:crossed}
\end{table}

The arc-wise searches of bnlearn in Table~\ref{tab:learners} also score the joint distribution of the lagged
variables. With a separable Gaussian score and a blacklist of arcs into lagged variables, that score splits
into a lag-block term and a term for the current variables given the lags, and the lag-block term does not
change the global optimum of the current part; the arc-wise searches, however, may follow different paths.
We therefore repeated the comparison with the score, the permitted arcs and the treatment of the lagged
variables held fixed, implementing hill climbing and tabu search on our conditional score
(Table~\ref{tab:crossed}; tabu search was within two errors of hill climbing in every cell). From seven to
thirty variables the search over orderings made 8 to 28 fewer errors than hill climbing restarted for the
same computing time, under either penalty, the paired differences being three to six standard errors. The
arc-wise searches stopped at scores well above that of the true graph, increasingly so as $n$ grew, and at
$k=20$ their errors hardly fell from $n=1000$ to 3000 (from 40.7 to 38.4 with the BIC in a single run),
whereas those of the search over orderings halved or better. From $k=20$ their errors were mostly false
inclusions; our interpretation, which we did not check on the move histories, is that once a proxy arc is in,
single-arc moves do not remove it.

The advantage is one of optimisation efficiency, not of a different target. With a hundred times the
computing time, restarted hill climbing approached the full search in R7 (9.9 against 7.4 errors at $n=500$
and 3.7 against 1.8 at $n=2000$ with the BIC). At $k=50$, where a fit takes several minutes, the comparison
with the local search over orderings with three restarts is inconclusive: 14.4 against 11.7 errors, a paired
difference of 2.75 in favour of the search over orderings with standard error 2.28 over twelve designs, while
hill climbing reached the lower score in eight of them. The penalty acts separately: the
extended BIC improved both kinds of search from $k=20$.

\subsection{Larger systems and the penalty}\label{sec:large}

\begin{table}[t]\centering\small
\begin{tabular}{rrrrrr}
\toprule
$k$ & $n$ & Designs & BIC & Ext.\ BIC & Ext.\ BIC $-$ BIC (s.e.)\\
\midrule
20 & 300 & 18 & 36.3 & 22.3 & $-14.1$ (1.9)\\
 & 1000 & 18 & 13.2 & 4.7 & $-8.5$ (0.9)\\
 & 3000 & 18 & 5.7 & 1.0 & $-4.7$ (0.6)\\
25 & 300 & 12 & 57.3 & 27.3 & $-30.0$ (2.4)\\
 & 1000 & 12 & 19.8 & 6.6 & $-13.3$ (1.2)\\
 & 3000 & 12 & 7.5 & 1.0 & $-6.5$ (0.8)\\
30 & 300 & 6 & 78.7 & 37.8 & $-40.8$ (3.8)\\
 & 1000 & 6 & 26.8 & 6.3 & $-20.5$ (2.6)\\
 & 3000 & 6 & 12.0 & 1.5 & $-10.5$ (1.4)\\
50 & 1000 & 12 & 69.7 & 11.7 & $-58.0$ (3.2)\\
\bottomrule
\end{tabular}
\caption{BIC and extended BIC ($\gamma=0.5$) in larger systems: mean
skeleton errors, one series per design, local search with three restarts (version 0.5.0 of the software). Last column: mean paired difference
with its standard error across designs.}\label{tab:large}
\end{table}

With the BIC the errors grow quickly with the number of variables (Table~\ref{tab:large}), and the reason is
the criterion, not the optimiser. At $k=30$ the search with three restarts found a DAG with a lower BIC than the true one in
all 18 design--size cells, by 103, 32 and 18 units on average at $n=300$, 1000 and 3000, so in these samples the
true class is not the BIC minimiser and no improvement of the optimiser could make a BIC search return it.
The mechanism is the one of Proposition~\ref{prop:grow}(c): on the same series, adding a single null lagged
variable to a true parent set lowered the BIC at 17, 10 and 7 of the 30 nodes on average at the three
sample sizes, and did so in every cell. These finite-sample checks do not verify the asymptotic assumptions
of the proposition. Without restarts Table~\ref{tab:large} is much the same (at
$k=50$, 72.1 and 13.2 errors).

The extended BIC addresses the cause. It made fewer errors than the BIC in every design at $k=25$, 30 and 50
and in 17 or 18 of the 18 designs at each size at $k=20$; the mean errors fell by 39 to 88\,\%. The gain is in false inclusions: at $k=50$ the extended BIC misses 7.2 true lagged links against
4.6 for the BIC but includes 3.4 spurious ones against 50.3, and makes 1.1 contemporaneous errors against
14.8. At $k=15$ the same holds (3.4 against 7.2 errors at $n=1000$, and class recovery 0.88 against 0.30 at
$n=10\,000$). An adaptive lasso screen placed before the BIC search removes some of the spurious
candidates but recovers only part of the gain (at $k=50$, 41.8 errors against 69.7 for the BIC alone). Among the values tested, the choice $\gamma=0.5$ is not critical (Table~\ref{tab:gamma}): each tested
$\gamma>0$ reduced the mean errors of the BIC from $k=15$, the errors for $\gamma=0.5$, 0.75 and 1 were
similar in the larger designs, $\gamma=0.25$ left some overfitting, and larger values cost accuracy clearly
only in R7, where many true links are weak.

\begin{table}[t]\centering\small
\begin{tabular}{llrrrrrr}
\toprule
Design & Measure & $n$ & $\gamma=0$ (BIC) & 0.25 & 0.5 & 0.75 & 1\\
\midrule
Model 1 & class recovery & 100 & 0.52 & 0.65 & 0.74 & 0.79 & 0.81\\
 & & 500 & 0.84 & 0.91 & 0.94 & 0.96 & 0.97\\
R7 & errors & 500 & 8.6 & 8.4 & 8.8 & 9.4 & 10.2\\
 & & 2000 & 1.3 & 1.0 & 0.9 & 1.1 & 1.3\\
$k=15$ & errors & 1000 & 7.2 & 4.2 & 3.4 & 3.5 & 3.8\\
$k=20$ & errors & 1000 & 13.2 & 6.9 & 4.7 & 4.3 & 4.5\\
$k=25$ & errors & 1000 & 19.8 & 8.6 & 6.6 & 6.2 & 6.3\\
$k=30$ & errors & 1000 & 26.8 & 9.3 & 6.3 & 6.5 & 6.7\\
$k=50$ & errors & 1000 & 69.7 & 20.8 & 11.7 & 10.7 & 12.5\\
\bottomrule
\end{tabular}
\caption{Sensitivity to the extended-BIC parameter $\gamma$: share of runs recovering the true class
(Model~1) or mean skeleton errors (other designs) of the unrestricted search. Replications as in Tables~1
and~\ref{tab:large}, 60 for $k=15$; for $k\ge20$ the series and the search (three restarts) are those of
Table~\ref{tab:large}. All sample sizes are in the Supplement.}\label{tab:gamma}
\end{table}

\subsection{When identification through lags fails}\label{sec:lags}

\begin{table}[t]\centering\small\setlength{\tabcolsep}{4pt}
\begin{tabular}{rrrrrrrrrr}
\toprule
Non-zero & Arrows & \multicolumn{2}{c}{Lagged links} & \multicolumn{2}{c}{Truly compelled} &
\multicolumn{2}{c}{Reported arrows} & \multicolumn{2}{c}{Skeleton}\\
cross-lag & compelled & \multicolumn{2}{c}{missed} & \multicolumn{2}{c}{arrows wrong} &
\multicolumn{2}{c}{correct} & \multicolumn{2}{c}{errors}\\
cells (of 42) & (of 150) & $n=500$ & 2000 & $n=500$ & 2000 & $n=500$ & 2000 & $n=500$ & 2000\\
\midrule
0 & 150 & 0.0 & 0.0 & 0\,\% & 0\,\% & 99\,\% & 99\,\% & 0.4 & 0.2\\
7 & 150 & 0.5 & 0.0 & 2\,\% & 0\,\% & 97\,\% & 100\,\% & 1.0 & 0.2\\
14 & 150 & 1.9 & 0.1 & 6\,\% & 0\,\% & 92\,\% & 99\,\% & 2.9 & 0.3\\
21 & 149 & 5.0 & 0.4 & 18\,\% & 2\,\% & 78\,\% & 97\,\% & 7.0 & 0.7\\
28 & 148 & 9.2 & 2.0 & 41\,\% & 13\,\% & 49\,\% & 84\,\% & 12.6 & 3.3\\
35 & 141 & 14.4 & 6.4 & 45\,\% & 34\,\% & 43\,\% & 54\,\% & 17.7 & 8.9\\
42 & 42 & 18.3 & 10.8 & 65\,\% & 48\,\% & 8\,\% & 13\,\% & 21.4 & 12.9\\
\bottomrule
\end{tabular}
\caption{Lag density and orientation: the contemporaneous graphs of R7 with one lag, an autoregressive
diagonal and a growing number of non-zero off-diagonal cells in the lag-1 matrix. Full search, extended BIC,
100 series per cell; lagged links missed and skeleton errors are means per fit. ``Truly compelled arrows
wrong'': share of the arrows compelled in truth that the fit compels in the opposite direction. ``Reported arrows
correct'': share of the arrows the fit reports as compelled that are compelled in that direction in truth
(totals over all fits).}\label{tab:lags}
\end{table}

The designs so far are favourable to identification through lags: their lag matrices are sparse and all 150
contemporaneous arrows of R7 are compelled. To see how far this carries, we kept the contemporaneous graphs
of R7 and replaced the lag structure by one lag with an autoregressive diagonal and a growing number of
non-zero cross-lag coefficients (Table~\ref{tab:lags}).

Orientation degrades gradually, and well before the lags stop carrying information. Up to a third of the
cells non-zero, at most 6\,\% of the arrows are oriented wrongly at $n=500$ and none at $n=2000$. These
designs have one true lag and every method is fitted with $p=3$. R7 itself has 12 non-zero off-diagonal
cells in its first lag matrix but also links at lags 2 and 3, so Table~\ref{tab:lags} does not place it
exactly; at $n=500$ the extended-BIC search oriented 5\,\% of its arrows wrongly (Supplement). With 35 of the 42 cells non-zero the true
class still compels 141 of the 150 arrows, yet the fit orients 45\,\% of them wrongly at $n=500$ and
34\,\% at $n=2000$. The errors follow the number of true lagged links that the fit omits: a false
exclusion can leave a triple unshielded and so create a v-structure, and with it an arrow that the data do not
support. A larger sample moves
the problem to higher densities but does not remove it: when the lag-1 matrix is full, so that the lags carry
no orientation information, the fit still reports more than 90\,\% of the unidentified edges as compelled.

The reader of a fitted graph faces the inverse question: of the arrows it reports as compelled, how many are
compelled in that direction in truth? Up to a third of the cells non-zero, 92 to 100\,\% are
(Table~\ref{tab:lags}). At 28 non-zero cells and $n=500$ the true classes still compel 148 of the 150
arrows, yet only 49\,\% of the arrows the fit reports as compelled are compelled in the reported direction:
the directions can be unreliable even where almost every true direction is identifiable. With 35 non-zero
cells about half are correct, and with a full lag matrix only 8 and 13\,\%
at $n=500$ and 2000: the fit reports more than eight compelled arrows per graph when the truth has about two,
and three quarters of what it reports lie on links that are reversible in truth or absent.

A search on the residuals of the full VAR uses no lag exclusion, and its population graph is the same at
every density. It orients far fewer arrows than our search, but once half to two thirds of the cells are non-zero it makes fewer wrong
orientations. Weak dynamics have the same effect as density, whereas unequal innovation variances, strong
autoregression or no cross-lags at all leave orientation as reliable as in R7; and with no dynamics at all a
few spurious lagged links are enough for 51 and 38\,\% of the unidentified edges to be reported as compelled at
$n=500$ and 2000, against 39 and 29\,\% for the search on the residuals, which can also change the
contemporaneous skeleton or its colliders (Supplement): both mechanisms contribute to false orientations.

The practical consequence is that an arrow should be reported with what identifies it. The software marks
each compelled arrow as identified by the contemporaneous graph alone or only through lag exclusions. An orientation identified through lag exclusions can survive some violations of those
exclusions; Section~\ref{sec:macro} examines which additions would make each fitted arrow reversible.

\subsection{Lag order}
With the lag order chosen on the reduced-form VAR, with orders up to the true one plus three compared on a common sample,
the Akaike information criterion (AIC) performed well in our designs: in R7 at $n=500$ the search with the
extended BIC made 8.8 errors with the AIC order against 8.6 with the true one, but 23.5 with the BIC order,
which underfitted in every run; at $k=15$ the BIC chose $p=1$ every time and the errors rose from 3.2 to 23.9.
An under-fitted order is far more costly than a slightly over-fitted one, and our other conclusions assume a
correct or generous order. Propositions~\ref{prop:min} and~\ref{prop:be} fix the lag order; extending them
to an estimated order would need an argument for the selection step. With a hundred variables the unrestricted search takes close to an hour and a screen becomes a
computational device, which we do not pursue here.

%=====================================================================
\section{Illustration: six US macroeconomic series}\label{sec:macro}

\begin{figure}[t]\centering
\begin{minipage}[c]{0.45\textwidth}\includegraphics[width=\textwidth]{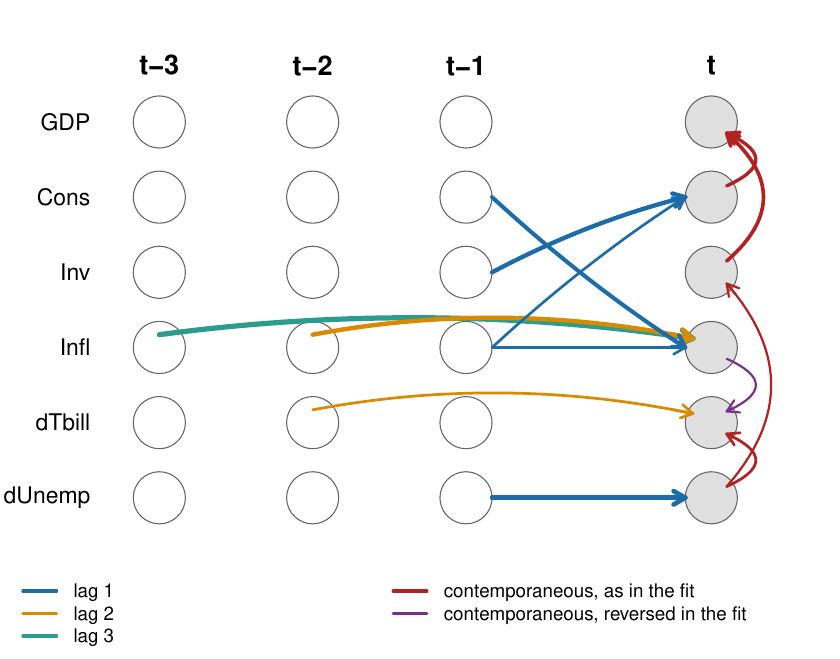}\end{minipage}\hfill
\begin{minipage}[c]{0.53\textwidth}\footnotesize\centering\setlength{\tabcolsep}{3.5pt}
\begin{tabular}{lrrrl}
\toprule
Arrow & coef. & adj. & arrow & identified by\\
\midrule
Cons $\to$ GDP & 0.59 & 1.00 & 0.72 & contemp.\ graph\\
Inv $\to$ GDP & 0.13 & 1.00 & 0.76 & contemp.\ graph\\
dUnemp $\to$ dTbill & $-1.01$ & 0.95 & 0.62 & lag exclusions\\
dUnemp $\to$ Inv & $-8.71$ & 0.95 & 0.55 & lag exclusions\\
dTbill $\to$ Infl & 0.21 & 0.88 & 0.34 & lag exclusions\\
dUnemp $\to$ Cons & $-1.10$ & 0.81 & 0.35 & lag exclusions\\
\bottomrule
\end{tabular}
\end{minipage}
\caption{The fit to six US series ($p=3$, extended BIC). Left: lagged links present in more than 50\,\% of
2000 moving-block bootstrap resamples of the window rows (blocks of 8), and contemporaneous arrows compelled in
that direction in the fitted class of more than 50\,\% of them; ``as in the fit'' and ``reversed in the fit''
compare the direction with the fit to the data; line width increases with the frequency. Right: the six
contemporaneous links of the fit, with coefficients $\beta_{ij}$, the share of all resamples whose fitted graph
contains the link in either direction (adj.), the share whose fitted class compels it in the fitted direction
(arrow), and what identifies the orientation.}\label{fig:macro}
\end{figure}

We analyse the quarterly \texttt{macrodata} set distributed with statsmodels \citep{seabold2010}, 1959Q2 to
2009Q3 ($n=202$): 100 times the quarterly log differences of real GDP, real consumption, real investment and
the consumer price index (inflation, not annualised), and the first differences, in percentage points, of
the three-month Treasury-bill rate and of the unemployment rate. The AIC, comparing orders 0 to 6 on the
common sample of $n-6$ observations, chooses $p=3$, so $N=199$; the search is the full one with the
extended BIC, and the bootstrap keeps $p=3$ fixed (Supplement). The fitted VAR is stable, with companion
spectral radius 0.868. The fit has six contemporaneous links, eight cross-lagged links and
six autoregressive terms (Figure~\ref{fig:macro}). GDP growth is a collider of consumption and investment
growth, with coefficients 0.59 and 0.13; these two series are components of GDP, so the collider reflects the
accounting relation among the three and is not independent evidence that structural shocks have been
identified. The other arrows run from changes in unemployment to consumption, investment and the bill rate,
and from the bill rate to inflation.

All six arrows are compelled given the time order, but only the two into GDP are identified by the
contemporaneous graph alone; the other four rest on lag exclusions. The bootstrap separates the two groups in
the same way. All six adjacencies appear in at least 81\,\% of the resamples and no excluded pair in more than
47\,\%. The arrows into GDP keep their direction in 72 and 76\,\% of the resamples; of the four that depend on
lag exclusions, two are kept in 55 and 62\,\% and two are more often reversed than kept. On series simulated
from the fitted model the fit oriented each arrow correctly in 88 to 98\,\% of cases, while the same bootstrap
kept it in 64 to 83\,\% of the resamples, so in this model bootstrap frequencies understate reliability; we do
not claim that the same holds for the economy that generated the data. The structural residuals are not
Gaussian: their sample kurtosis is 12.3 for the bill-rate change and 7.7 for inflation, against 3, and the
lag-one autocorrelations of their squares are 0.53 and 0.40, which limits how far the Gaussian calibration
extends to the observed series. Against that benchmark
the two arrows into GDP are about as stable as the fitted model implies (0.74 against 0.82 on average) and
the four lag-dependent arrows are not (0.46 against 0.68). The skeleton and the GDP collider look trustworthy;
the remaining arrows are best read as hypotheses.

Which exclusions hold the four up? A link can be reversible only if its two ends have the same lagged
parents (Section~\ref{sec:equiv}), so we added lagged edges to the fitted DAG, keeping its contemporaneous
directions, and recomputed the class. No single added edge, among the 94 absent lag-to-current links, makes
any of the six arrows reversible. Unemployment $\to$ consumption, $\to$ investment and $\to$ bill rate each
become reversible when four absent lagged links are added, those that give the two ends the same lagged
parents; for consumption, the first lags of inflation and of the bill rate into unemployment and the first
lags of consumption and unemployment into consumption. By the rule of Section~\ref{sec:equiv} no addition of
fewer edges can do so. For bill rate $\to$ inflation the endpoints are not enough: reversing it while keeping
unemployment $\to$ bill rate would create the new unshielded collider unemployment $\to$ bill rate
$\leftarrow$ inflation, so that arrow must reverse too, and unemployment, the bill rate and inflation must all
have the same lagged parents. Each must acquire every lagged parent present at any of the three, which is
sixteen additions. Attainability was checked with explicit witnesses: after the four additions, the DAG with
consumption, investment or the bill rate respectively pointing to unemployment, and after the sixteen, the
DAG with inflation $\to$ bill rate $\to$ unemployment, all other edges unchanged, are acyclic and have the
skeleton and v-structures of the augmented graph before the reversal. So four and sixteen are the minima over
all additions. The
lag-dependent arrows therefore rest on several exclusions jointly, none decisive alone.

What does the bootstrap vary? We drew a second set of 2000 window-row resamples (blocks of 8), so the full-refit
frequencies below differ slightly from those of Figure~\ref{fig:macro}, and applied every procedure to the
same draws; ``kept'' means compelled, in the equivalence class, in the fitted direction, and the lag order is
fixed at $p=3$. First, we held each resample's entire fitted contemporaneous DAG fixed, replaced its lagged
edges by those of the original fit and recomputed only the class: all six links had the same classification
as in the full refit in every resample (kept, reversed, unoriented and absent coded separately). Second, we
held each resample's contemporaneous skeleton fixed and reselected its directions, minimising the extended
BIC over all acyclic orientations twice, once with the lagged edges selected in that
resample and once with those of the original fit, so that data, skeleton and optimiser are the same and only
the lagged edges differ; no minimum was tied. With the resample's own lagged edges the exhaustive search
reproduced the full refit's classification of all six links in 99\,\% of the resamples. With the original
fit's lagged edges the fitted directions were kept far more often: 0.87 and 0.90 for the arrows into GDP
(0.72 and 0.77 with the resample's own), 0.59, 0.91 and 0.71 for unemployment $\to$ consumption, investment
and bill rate (0.35, 0.52, 0.63), and 0.50 for bill rate $\to$ inflation (0.34); the two reoptimised
classifications of the six links agreed in only 16\,\% of the resamples. Within this exercise, with the lag
order and the skeleton fixed, the difference is due to the lagged edges alone: the directions the
search selects depend on the lagged links selected with them, and much of the instability of the
orientations comes from the selection of lags in each resample, not from the contemporaneous skeleton alone.

%=====================================================================
\section{Discussion}\label{sec:discussion}

The lagged predictors of a structural VAR can distinguish contemporaneous directions that its reduced-form
innovations leave unresolved. A penalised-likelihood search over contemporaneous orderings, with all lagged
variables as eligible parents, exploits this. For a fixed number of variables and a fixed lag order, and under faithfulness given the past, the
penalised minimiser and the dynamic programme with backward elimination that the software uses for up to ten
variables recover the equivalence class under the time ordering with probability tending to one. In our designs the search had fewer mean skeleton errors than the published learners of
Table~\ref{tab:learners}, except in a three-variable model; TS-BOSS, given the same score, returned similar graphs, with differences of at most
1.6 errors in either direction; and with the score held fixed our search was more accurate than arc-wise search at equal computing time for up to thirty
variables, the comparison at fifty being inconclusive (a paired difference of 2.75 errors, standard error
2.28). The penalty matters as much as the search: the BIC overfits as the
number of candidates grows, as Proposition~\ref{prop:grow} predicts under its conditions when nearly
uncorrelated null candidates grow like $n^{\kappa'}$ with $\kappa'>1/2$, and the extended BIC removes most of
the excess, with mean skeleton errors close to those of the BIC in R7 and better class recovery in Model~1 at
small $n$.

The graph describes the contemporaneous structure under assumptions that the data cannot check: the effects
are acyclic within a period, there are no omitted common causes, and the distribution is faithful to the
augmented window graph given the past (Proposition~\ref{prop:min}). A compelled arrow identifies a direction;
an invariant parent set identifies the structural equation and its shock (Section~\ref{sec:equiv}). With
Gaussian innovations only the equivalence class is identified, so a structural reading of an arrow needs it
to be compelled, and the compelled status of an estimated arrow is conditional on the estimated lag
exclusions. Standard errors and impulse responses computed from the selected graph are conditional on it;
we give no inference that accounts for the selection, and the bootstrap of Section~\ref{sec:macro} measures
stability only. The DAG is a transparent way to extract the restrictions that the data support and to show where
they stop; it does not replace economic arguments, instruments, long-run or sign restrictions, or
non-Gaussian identification \citep{kilian2017,hyvarinen2010}.

\emph{Limitations.} (i) The local search over orderings is not covered by the theory; it is validated against
the full search only for $k\le15$, it needed restarts at $k=50$, and there a restarted arc-wise search given
the same time reached a lower score in eight of twelve designs. (ii) For a growing number of variables the
theory covers parent selection for a given ordering, with bounded parent sets and, for its negative part, an
assumed Gaussian approximation. (iii) Orientation through lag exclusions is not robust to dense or weak lag
structure (Section~\ref{sec:lags}); in those cases the output overstates how many arrows are identified.
(iv) The designs are Gaussian, acyclic and causally sufficient with stationary dynamics: apart from the lag
order and the lag structure, we did not study how the method fails under omitted common causes,
contemporaneous feedback or non-Gaussian innovations. There is one series per design for
$k\ge20$, and the comparisons with the published learners of Table~\ref{tab:learners} use 20 series or fewer in one
configuration each.

\emph{Software.} The R package \texttt{gmsvar} (version 0.5.0,
\url{https://github.com/MarcoNZreal/gmsvar}) chooses the lag order, runs the search with the extended BIC, and returns the estimated SVAR, the equivalence
class with what identifies each arrow, impulse responses and a moving-block bootstrap. The simulation scripts
and the data, with the result files behind every table and figure, are archived at \url{https://github.com/MarcoNZreal/gmsvar-replication}.

\emph{Data availability.} The six US macroeconomic series of Section~\ref{sec:macro} are the quarterly
\texttt{macrodata} set distributed with statsmodels \citep{seabold2010}, compiled from the Federal Reserve
Economic Data and the Bureau of Labor Statistics and in the public domain; the transformed series are
supplied as supporting information. The R package \texttt{gmsvar}, version 0.5.0, is available at
\url{https://github.com/MarcoNZreal/gmsvar}, and the scripts, data and result files that reproduce every
table and figure of the paper and its supplement at \url{https://github.com/MarcoNZreal/gmsvar-replication}.

%=====================================================================
\section{Acknowledgements}\label{sec:acknowledgements}
During the preparation of this manuscript, the authors used ChatGPT (Astra 6.2 and Sol 6) and Claude (Sonnet 5.5 and Opus 5.5) to assist with brainstorming, suggesting mathematical constructions and proof strategies and checking mathematical proofs; improving the clarity and grammar of the exposition; designing and coding simulations and literature search. After using these tools, the authors thoroughly verified every mathematical argument, proof, piece of simulation code, and revised passage, and take full responsibility for the correctness, originality and intellectual integrity of the content.

%=====================================================================
\appendix
\section{Proofs of Propositions 1 and 2}

\subsection*{Proof of Proposition~\ref{prop:min}}
\emph{Notation.} Let $\mathcal G$ be the finite set of SVAR DAGs. For $G\in\mathcal G$ with parent sets $P_i$
let $G^+$ be its augmented graph and $M(G)$ the set of positive definite $m\times m$ matrices that are
covariance matrices of a Gaussian vector Markov to $G^+$; the joint distribution of the lagged variables is unrestricted in $M(G)$. Let $\Sh$
be the sample covariance of the window and $\Sigma$ the true one, and $|G|=\sum_{i\le k}|P_i|$. For $S$
positive definite and index sets $A,B$ let $S_{AB}$ be the submatrix with rows $A$ and columns $B$
(singletons written as indices, so that $S_{LL}$ is the lag block), let
$\hat\sigma_i^2(P;S)=S_{ii}-S_{iP}S_{PP}^{-1}S_{Pi}$, so that $\sigma_i^2(P)=\hat\sigma_i^2(P;\Sigma)$ is the
population residual variance of Section~\ref{sec:method}, and
\[
T_G(S)=\sum_{i\le k}\log\hat\sigma_i^2(P_i;S),\qquad \mathrm{IC}(G)=N\,T_G(\Sh)+|G|\lambda_N .
\]
$\Sigma$ is positive definite and $\Sigma\in M(\Gs)$. The two facts used about the data are $\Sh\to\Sigma$ in
probability and $\sqrt N(\Sh-\Sigma)=O_p(1)$ \citep{lutkepohl2005}. All statements are made on the event that
$\Sh$ is positive definite, whose probability tends to one.

\emph{Step 1: maximum-likelihood fit.} For $S$ positive definite let $\Sh_G(S)\in M(G)$ be the Gaussian
maximum-likelihood fit to $S$: each vertex of $G^+$ is regressed on its parents, the lagged vertices on the
lagged vertices of lower index, so that this part fits $S_{LL}$ exactly. Then
$\log\det\Sh_G(S)=T_G(S)+\log\det S_{LL}$ and $\operatorname{tr}\{\Sh_G(S)^{-1}S\}=m$. Define
\[
\delta_G(S)=\operatorname{tr}\{\Sh_G^{-1}S\}-m-\log\det(\Sh_G^{-1}S)=\log\det\Sh_G(S)-\log\det S\ \ge0,
\]
the inequality being $x-1-\log x\ge0$ applied to the eigenvalues. Hence
\begin{equation}\label{eq:T}
T_G(S)=\delta_G(S)+\log\det S-\log\det S_{LL}.
\end{equation}

\emph{Step 2: models that contain the truth.} Suppose $\Sigma\in M(G)$. Because $\Sh_G(S)$ minimises
$\Sigma'\mapsto\log\det\Sigma'+\operatorname{tr}(\Sigma'^{-1}S)$ over $M(G)$,
$0\le\delta_G(S)\le\delta(S,\Sigma):=\operatorname{tr}(\Sigma^{-1}S)-m-\log\det(\Sigma^{-1}S)$. If
$\lambda_1,\dots,\lambda_m$ are the eigenvalues of $\Sigma^{-1/2}S\Sigma^{-1/2}$, then
$\delta(S,\Sigma)=\sum_j(\lambda_j-1-\log\lambda_j)\le\sum_j(\lambda_j-1)^2$ as soon as all
$|\lambda_j-1|\le1/2$. With $S=\Sh$ the right-hand side is $O_p(N^{-1})$, so
\begin{equation}\label{eq:rate}
0\le N\delta_G(\Sh)\le N\delta(\Sh,\Sigma)=O_p(1)\qquad\text{for every $G$ with }\Sigma\in M(G).
\end{equation}

\emph{Step 3: models that do not contain the truth.} Put $c_G=\delta_G(\Sigma)\ge0$. Then $c_G=0$ exactly when
$\Sigma=\Sh_G(\Sigma)\in M(G)$, so $c_G>0$ if $\Sigma\notin M(G)$, and $c_{\Gs}=0$. The map $S\mapsto T_G(S)$
is continuous at $\Sigma$, so $T_G(\Sh)\to T_G(\Sigma)$ in probability, and by \eqref{eq:T}
$T_G(\Sigma)-T_{\Gs}(\Sigma)=c_G$.

\emph{Step 4: graph-theoretic facts under faithfulness.} Let $I=\{G:\Sigma\in M(G)\}$. (i) If $G\in I$ then
$|G|\ge|\Gs|$, with equality if and only if $G$ is equivalent to $\Gs$. (ii) If $G$ is equivalent to $\Gs$ then
$M(G)=M(\Gs)$ and $\mathrm{IC}(G)=\mathrm{IC}(\Gs)$ for every sample. Only (i) uses faithfulness, and the rest
of the proof uses only (i).

For (i), let $G\in I$. Let $a,b$ be non-adjacent in $G^+$ and not both lagged, and let $b$ be the later of the
two in a topological order of $G^+$ in which the lagged vertices come first; then $b$ is current, and
$S=\mathrm{nd}_{G^+}(b)\setminus\{a\}$, with $\mathrm{nd}$ the non-descendants of $b$ other than $b$, contains
$\pa_{G^+}(b)$ and every lagged vertex other than $a$. By the local Markov property and weak union,
$Z_a\indep Z_b\mid Z_S$, so by faithfulness given the past $a,b$ are d-separated by $S$ in $(\Gs)^+$. Suppose
$a,b$ are adjacent in $(\Gs)^+$ but not in $G^+$; this is then a contradiction. So the skeleton of $G^+$ contains
that of $(\Gs)^+$; every contemporaneous or lag-to-current edge appears in exactly one parent set, hence
$|G|\ge|\Gs|$, with equality if and only if the skeletons coincide. Assume that they do, and let $a-c-b$ be an
unshielded triple. Its endpoints are not both lagged, because the lagged vertices form a clique in both
augmented graphs; let $b$ be the later endpoint in $G^+$ and $S=\mathrm{nd}_{G^+}(b)\setminus\{a\}$ as above,
so that $a,b$ are d-separated by $S$ in $(\Gs)^+$. If $c$ is a collider in $(\Gs)^+$ but not in $G^+$, then the
non-collider $c$ is a parent of $b$ in $G^+$, since $a\leftarrow c\leftarrow b$ would place $b$ before $a$;
so $c\in S$, and conditioning on the collider $c$ opens
the path $a\to c\leftarrow b$ in $(\Gs)^+$, a contradiction. If $c$ is a collider in $G^+$ but not in
$(\Gs)^+$, then $c$ is a child of $b$ in $G^+$, so $c\notin S$, and the path through the non-collider $c$ is
open in $(\Gs)^+$, again a contradiction. Hence the graphs have the same skeleton and v-structures. For (ii), DAGs with the same skeleton and
v-structures have the same Gaussian Markov model \citep{verma1990,andersson1997}, so the fits coincide,
$T_G=T_{\Gs}$, and $|G|=|\Gs|$.

\emph{Step 5: comparison with the truth.} Let $G$ not be equivalent to $\Gs$. If $G\notin I$, then with
probability tending to one $N\{T_G(\Sh)-T_{\Gs}(\Sh)\}\ge Nc_G/2$ and
$\BIC(G)-\BIC(\Gs)\ge\tfrac12Nc_G-|\Gs|\lambda_N\to\infty$. If $G\in I$, then $|G|\ge|\Gs|+1$ and by
\eqref{eq:T} and \eqref{eq:rate}
$N\{T_G(\Sh)-T_{\Gs}(\Sh)\}=N\delta_G(\Sh)-N\delta_{\Gs}(\Sh)\ge-O_p(1)$, so
$\BIC(G)-\BIC(\Gs)\ge\lambda_N-O_p(1)\to\infty$. Since $\mathcal G$ is finite, the event
\[
B_N=\{\BIC(G)>\BIC(\Gs)\text{ for all $G\in\mathcal G$ not equivalent to }\Gs\}
\]
has $\Pp(B_N)\to1$.

\emph{Step 6: conclusion.} On $A_N=\{\pa^*(i)\subseteq\mathcal N_N(i)\ \forall i\}$ the structure $\Gs$ belongs
to $\mathcal S_N$. On $A_N\cap B_N$ every structure of $\mathcal S_N$ not equivalent to $\Gs$ has a larger
score than $\Gs$, so every minimiser over $\mathcal S_N$ is equivalent to $\Gs$, and
$\Pp(A_N\cap B_N)\to1$. \qed

\subsection*{Proof of Proposition~\ref{prop:be}}
Fix a topological ordering $\pi=(\pi_1,\dots,\pi_k)$ of the contemporaneous part of $\Gs$ and, for
$i=\pi_r$, let $C^\pi_i=\{\pi_1,\dots,\pi_{r-1}\}\cup L$ and $S_i=\pa^*(i)\subseteq C^\pi_i$.

\emph{Step 1: population facts.} By \eqref{eq:svar},
$Z_{i,t}=\sum_{j\in S_i}\beta_{ij}Z_{j,t}+\varepsilon_{i,t}$ with all $\beta_{ij}\ne0$. Each $Z_{j,t}$,
$j\in C^\pi_i$, is a function of structural shocks dated before $t$ and of the date-$t$ shocks of variables
that precede $i$ in $\pi$; since $D$ is diagonal and the process is Gaussian, $\varepsilon_{i,t}$ is
independent of $Z_{C^\pi_i,t}$. Hence, for every $P$ with $S_i\subseteq P\subseteq C^\pi_i$, the population
regression of $Z_i$ on $Z_P$ has coefficients $\beta_{ij}$ on $S_i$ and zero on $P\setminus S_i$, with
residual variance $\sigma_i^2$, and for $j\in S_i$
\[
\sigma_i^2(P\setminus\{j\})=\sigma_i^2+\beta_{ij}^2\operatorname{Var}\bigl(Z_j\mid Z_{P\setminus\{j\}}\bigr)>\sigma_i^2,
\]
because $\Sigma$ is positive definite. Let $a>0$ be the minimum of
$\log\sigma_i^2(P\setminus\{j\})-\log\sigma_i^2$ over the finitely many such triples $(i,P,j)$. If all true
parent sets are empty, take $a=1$, since assertion (i) of Step~2 is then vacuous.

\emph{Step 2: single deletions in the sample.} Let $D_N$ be the event that, for every $i\le k$ and every $P$
with $S_i\subseteq P\subseteq C^\pi_i$, (i) $s_i(P\setminus\{j\})>s_i(P)$ for all $j\in S_i$ and
(ii) $s_i(P\setminus\{j\})<s_i(P)$ for all $j\in P\setminus S_i$. For (i), $\hat\sigma_i^2(\cdot)$ is a
continuous function of $\Sh$, so $N\{\log\hat\sigma_i^2(P\setminus\{j\})-\log\hat\sigma_i^2(P)\}\ge Na/2$ with
probability tending to one and $s_i(P\setminus\{j\})-s_i(P)\ge Na/2-\lambda_N\to\infty$. For (ii),
$N\{\log\hat\sigma_i^2(P\setminus\{j\})-\log\hat\sigma_i^2(P)\}=-N\log(1-\hat r^2)$ with $\hat r$ the sample
partial correlation of $Z_i$ and $Z_j$ given $Z_{P\setminus\{j\}}$, whose population value is zero by Step~1;
$\sqrt N\hat r=O_p(1)$, so $s_i(P\setminus\{j\})-s_i(P)=O_p(1)-\lambda_N\to-\infty$. There are finitely many
triples, so $\Pp(D_N)\to1$.

\emph{Step 3: backward elimination along $\pi$.} On $D_N$, start backward elimination for node $i$ from any
$C$ with $S_i\subseteq C\subseteq C^\pi_i$. While the current set $P$ contains $S_i$ strictly, by (ii) some
deletion lowers the score and by (i) every deletion that lowers it removes an element of $P\setminus S_i$, so
the next set again contains $S_i$. When $P=S_i$, by (i) no deletion lowers the score. Hence the procedure
returns $S_i$.

\emph{Step 4: the dynamic programme.} Let $\widehat{\mathrm{best}}(U)$ be the quantity computed by
Algorithm~1. The structure $\hat{\mathcal G}_{\mathrm{BE}}$ obtained by tracing back the minimisers is an SVAR
DAG in $\mathcal S_N$, and by decomposability
$\BIC(\hat{\mathcal G}_{\mathrm{BE}})=\widehat{\mathrm{best}}(\{1,\dots,k\})$. On $A_N\cap D_N$ the candidate
set of $\pi_r$ given $U_{r-1}=\{\pi_1,\dots,\pi_{r-1}\}$ contains $S_{\pi_r}$ and is contained in
$C^\pi_{\pi_r}$, so by Step~3 and induction on $r$
\[
\widehat{\mathrm{best}}(U_r)\le\widehat{\mathrm{best}}(U_{r-1})+s_{\pi_r}(S_{\pi_r})\le\sum_{l\le r}s_{\pi_l}(S_{\pi_l}),
\qquad\text{so}\qquad\BIC(\hat{\mathcal G}_{\mathrm{BE}})\le\BIC(\Gs).
\]

\emph{Step 5: conclusion.} On $B_N$ every structure not equivalent to $\Gs$ has a score strictly larger than
that of $\Gs$. Hence on $A_N\cap B_N\cap D_N$, whose probability tends to one,
$\hat{\mathcal G}_{\mathrm{BE}}$ is equivalent to $\Gs$. \qed

%=====================================================================

%=====================================================================
\clearpage
\setcounter{section}{0}\setcounter{table}{0}\setcounter{equation}{0}\setcounter{figure}{0}
\renewcommand{\thesection}{S\arabic{section}}
\renewcommand{\thetable}{S\arabic{table}}
\renewcommand{\theequation}{S\arabic{equation}}
\renewcommand{\thefigure}{S\arabic{figure}}
\begin{center}{\Large Supplementary material}\end{center}
\addcontentsline{toc}{section}{Supplementary material}

\noindent References to sections, propositions, tables and equations without the prefix S are to the main
paper. Section~\ref{S-sec:proof} proves Proposition~3. Section~\ref{S-sec:index} indexes the simulation studies.
Sections~\ref{S-sec:small}--\ref{S-sec:app} give the details behind Sections~6 and~7.

%=====================================================================
\section{Proof of Proposition 3}\label{S-sec:proof}

The notation is that of Section~4.2: $C_i$, $S_i$, $\beta_{ij}$, $\hat S_i$, the penalty
$\lambda_N=\log N+2\gamma\log(m-1)$ and assumptions (G1)--(G3). Assumption (GA), used in part (c) only, is
stated at the start of the proof of that part.

\begin{proof}[Proof of Proposition 3]
Fix a current variable $i$ and write $C=C_i$, $S=S_i$, $s=|S|$, $q=|C|\le m-1$, $\varepsilon_t=\varepsilon_{i,t}$
and $\sigma^2=\sigma_i^2$; constants $c,C_0,C_1,\dots$ depend only on $\underline\lambda$, $\bar\lambda$, the bound
on the spectral density, $K$ and $\bar s$, so that all bounds are uniform in $i$. Let $\Ft$ be the $\sigma$-field
generated by all structural shocks dated before $t$ and by the date-$t$ shocks of the variables that precede $i$
in the ordering. Every candidate $Z_{j,t}$, $j\in C$, is $\Ft$-measurable, $\varepsilon_t$ is independent of
$\Ft$ and $\mathcal F_{t+1}$-measurable, and $Z_{i,t}=Z_{S,t}'\beta+\varepsilon_t$.

\emph{Intercept.} The scores are computed from the centred sample covariance, that is, every regression
contains an intercept. We therefore adjoin the constant $Z_{0,t}\equiv1$ to the window. It is
$\Ft$-measurable; it belongs to every conditioning set below without being counted in its size; and $\Sigma$,
$\Sh$ denote the second-moment matrices of $(1,Z_t')'$ and their sample versions (divisor $N$). Since
$\mathrm EZ_t=0$, $\Sigma=\mathrm{diag}(1,\operatorname{Var}Z_t)$, with eigenvalues in a fixed interval which we
still call $[\underline\lambda,\bar\lambda]$. With this convention the residual sums of squares below are those
of the centred regressions, and all regressors remain predictable.

\emph{Step 0: the good event.} By the deviation bound of Basu and Michailidis (2015, Proposition~2.4) for stable
Gaussian processes with bounded spectral density, the Gaussian tail bound for the sample means (whose variances
are at most $2\pi$ times the bound on the spectral density, divided by $N$), and a union bound over the
$(m+1)^2$ entries, the event $E_N$ on which every entry of $\Sh$ differs from its population value by at most
$C_0\sqrt{\log m/N}$ has $\Pp(E_N^c)\le m^{-2}$. Since $\log m/N\to0$, on $E_N$ and for $N$ large every principal
submatrix of $\Sh$ of order at most $2K+2$ has its eigenvalues in $[\underline\lambda/2,2\bar\lambda]$. Fix
$\delta\in(0,1/8)$ and let
\[
X_N=\bigl\{(1-\delta)N\sigma^2\le\textstyle\sum_t\varepsilon_t^2\le(1+\delta)N\sigma^2\bigr\},\qquad
\Pp(X_N^c)\le2e^{-N\delta^2/8},
\]
by the chi-squared tail bound. The event $H_N$ is defined in Step~1; the good event is
$G_N=E_N\cap X_N\cap H_N$, and we show $\Pp(G_N^c)\le3m^{-2}$ for $N$ large.

\emph{Step 1: a sharp tail bound for scores.} Let $w_t$ be $\Ft$-measurable, $M_N=\sum_tw_t\varepsilon_t$ and
$V_N=\sum_tw_t^2$. Since $\varepsilon_t\mid\Ft\sim N(0,\sigma^2)$, $\exp(\theta M_N-\theta^2\sigma^2V_N/2)$ is a
martingale with mean one (Freedman, 1975), so
$\Pp\{M_N\ge\sigma\sqrt{xv},\,V_N\le v(1+\delta)\}\le\exp\{-x(1-\delta)/2\}$ (take $\theta=\sqrt{x/v}/\sigma$).
Covering $V_N\in[a,b]$ by $L=\lceil\log(b/a)/\log(1+\delta)\rceil$ intervals of ratio $1+\delta$ gives
\begin{equation}\label{S-eq:7}
\Pp\bigl(M_N^2\ge x\sigma^2V_N,\ a\le V_N\le b\bigr)\le2L\,e^{-(1-\delta)x/2}.
\end{equation}
For disjoint $D,P\subseteq C$ with $|D\cup P|\le2K$, let $W_t^\circ=Z_{D,t}-B'Z_{P,t}$ be the population residuals
of $Z_D$ on $Z_P$ (and the constant), $U=\sum_tW_t^\circ\varepsilon_t$,
$\mathcal W=\sum_tW_t^\circ W_t^{\circ\prime}$ and
\[
Q_{D\cdot P}=U'\mathcal W^{-1}U=\sup_{u\ne0}\frac{(u'U)^2}{u'\mathcal Wu}.
\]
$W_t^\circ$ is $\Ft$-measurable, and on $E_N$ the eigenvalues of $\mathcal W/N$ lie in a fixed interval
$[c_1,c_2]$. Let $\mathcal N_\delta$ be a $\delta$-net of the unit sphere of $\mathbb R^{|D|}$, of size at most
$(3/\delta)^{|D|}$. Comparing the maximiser with the nearest point of the net gives
$Q_{D\cdot P}\le Q^0(1+\delta\sqrt{c_2/c_1})/(1-\delta\sqrt{c_2/c_1})$, with $Q^0$ the maximum over the net, and
applying \eqref{S-eq:7} to $w_t=u_0'W_t^\circ$ for each $u_0\in\mathcal N_\delta$ (with $V_N\in[Nc_1,Nc_2]$), after
renaming $\delta$,
\begin{equation}\label{S-eq:8}
\Pp\bigl(Q_{D\cdot P}\ge\sigma^2x,\ E_N\bigr)\le C_1(\delta)\,e^{-(1-\delta)x/2}\qquad\text{for every }\delta\in(0,1),
\end{equation}
with $C_1(\delta)$ not depending on $N$, $D$, $P$ or $i$. Let $H_N$ be the event that
$Q_{D\cdot P}\le\sigma^2C_2\log m$ for all such pairs $(D,P)$ and that
$\max_{j\in C\cup\{0\}}|\sum_tZ_{j,t}\varepsilon_t|\le C_3\sqrt{N\log m}$. There are at most $m^{2K}$ pairs, so by
\eqref{S-eq:8} with $x=C_2\log m$, and \eqref{S-eq:7} with $w_t=Z_{j,t}$, $\Pp(H_N^c\cap E_N)\le m^{-2}$ for $C_2,C_3$
large. With Step~0, $\Pp(G_N^c)\le2m^{-2}+2e^{-N\delta^2/8}\le3m^{-2}$ for $N$ large, by (G2).

\emph{Lemma (sample residuals).} Let $\hat W_t=Z_{D,t}-\hat B'Z_{P,t}$ be the least-squares residuals of $Z_D$ on
$Z_P$ (and the constant), $\hat U=\sum_t\hat W_t\varepsilon_t$, $\hat{\mathcal W}=\sum_t\hat W_t\hat W_t'$ and
$\hat Q_{D\cdot P}=\hat U'\hat{\mathcal W}^{-1}\hat U$, which is the squared norm of the projection of
$\varepsilon$ on the sample residuals of $Z_D$ given $Z_P$. Then, on $G_N$ and for $N\ge N_0(\delta)$, for all
disjoint $D,P\subseteq C$ with $|D\cup P|\le2K$,
\begin{equation}\label{S-eq:lemma}
\hat Q_{D\cdot P}\le(1+2\delta)\,Q_{D\cdot P}+C_\delta\frac{(\log m)^2}{N}.
\end{equation}
\emph{Proof of the lemma.} On $E_N$, $\|\hat B-B\|\le C\sqrt{\log m/N}$, because $B$ and $\hat B$ are the same
smooth function of well-conditioned submatrices of bounded order of $\Sigma$ and $\Sh$. Hence, on $G_N$,
$\|\hat U-U\|=\|(\hat B-B)'\sum_tZ_{P,t}\varepsilon_t\|\le C\log m$. Since $\hat B$ minimises the residual sum
of squares, $\mathcal W-\hat{\mathcal W}=N(\hat B-B)'\Sh_{PP}(\hat B-B)$ is non-negative definite with norm at
most $C\log m$, so $\hat{\mathcal W}\ge(1-C\log m/(c_1N))\mathcal W$ and
$\hat{\mathcal W}^{-1}\le(1+\delta/2)\mathcal W^{-1}$ for $N$ large. Finally
$\hat U'\mathcal W^{-1}\hat U\le(1+\delta)U'\mathcal W^{-1}U+(1+1/\delta)(\hat U-U)'\mathcal W^{-1}(\hat U-U)
\le(1+\delta)Q_{D\cdot P}+(1+1/\delta)C^2(\log m)^2/(c_1N)$. Combining the two gives \eqref{S-eq:lemma}.
\hfill$\diamond$

\emph{Step 2: overfitting.} Let $\mathrm{RSS}(P)=N\hat\sigma_i^2(P)$ and $A\subseteq C\setminus S$ with
$|A|=j\le K-s$. Because $Z_i=Z_S'\beta+\varepsilon$,
$\mathrm{RSS}(S)-\mathrm{RSS}(S\cup A)=\hat Q_{A\cdot S}$ and
$\mathrm{RSS}(S\cup A)=\sum_t\varepsilon_t^2-N\bar\varepsilon^2-\hat Q_{(S\cup A)\cdot\emptyset}$, where
$\bar\varepsilon$ is the sample mean of the $\varepsilon_t$. On $G_N$, by \eqref{S-eq:lemma} and the definition of
$H_N$, $\hat Q_{(S\cup A)\cdot\emptyset}\le2\sigma^2C_2\log m$ and $N\bar\varepsilon^2\le C_3^2\log m$, so
$\mathrm{RSS}(S\cup A)\ge(1-2\delta)N\sigma^2$ for $N$ large. Since $\log(1+x)\le x$, $S\cup A$ beats $S$ only if
\[
j\lambda_N<N\log\frac{\mathrm{RSS}(S)}{\mathrm{RSS}(S\cup A)}\le\frac{N\hat Q_{A\cdot S}}{\mathrm{RSS}(S\cup A)}
\le\frac{\hat Q_{A\cdot S}}{(1-2\delta)\sigma^2},
\]
which by \eqref{S-eq:lemma} implies $Q_{A\cdot S}\ge(1-5\delta)\sigma^2j\lambda_N$ for $N$ large (the term
$(\log m)^2/N$ is $o(1)$). By \eqref{S-eq:8}, $\binom qj\le(m-1)^j$, $\lambda_N=\log N+2\gamma\log(m-1)$ and
$m-1\le cN^\kappa$,
\begin{align*}
\Pp(\exists A:\ S\cup A\text{ beats }S,\ G_N)
&\le\sum_{j=1}^{K}C_4(\delta)\,(m-1)^{j\{1-(1-6\delta)\gamma\}}N^{-(1-6\delta)j/2}\\
&\le C_5(\delta)\sum_{j\ge1}N^{j\{\kappa(1-\gamma)_+-1/2+C_6\delta\}},
\end{align*}
where $x_+=\max(x,0)$. This tends to zero, for $\delta$ small, when $\kappa(1-\gamma)<1/2$, that is
$\gamma>1-1/(2\kappa)$.

\emph{Step 3: underfitting.} Let $P\subseteq C$, $|P|\le K$, $D=S\setminus P\ne\emptyset$ and $T=P\cup S$, so
$|T|\le2K$. Then $\mathrm{RSS}(P)-\mathrm{RSS}(T)=\|\Pi(Z_D\beta_D+\varepsilon)\|^2$, where $\Pi$ is the projection
on the sample residuals of $Z_D$ given $Z_P$. On $G_N$,
$\|\Pi Z_D\beta_D\|^2=N\beta_D'\Sh_{DD\cdot P}\beta_D\ge N\underline\lambda b_N^2/2$, and by \eqref{S-eq:lemma}
$\|\Pi\varepsilon\|^2=\hat Q_{D\cdot P}\le2\sigma^2C_2\log m$. Hence
$\mathrm{RSS}(P)-\mathrm{RSS}(T)\ge(\sqrt{N\underline\lambda/2}\,b_N-\sigma\sqrt{2C_2\log m})^2\ge cNb_N^2$, because
$Nb_N^2/\log N\to\infty$ and $\log m=O(\log N)$. Also
$\mathrm{RSS}(S)-\mathrm{RSS}(T)=\hat Q_{(T\setminus S)\cdot S}\le2\sigma^2C_2\log m$ and
$\mathrm{RSS}(S)\le\sum_t\varepsilon_t^2\le2N\sigma^2$. Therefore, on $G_N$ and for $N$ large, the penalised
score of $P$ exceeds that of $S$ by at least
$N\log\{1+cb_N^2/(4\sigma^2)\}-K\lambda_N\ge c'\min(Nb_N^2,N)-K\lambda_N>0$, uniformly in $P$ and $i$.

\emph{Conclusion.} Any $P\ne S$ with $|P|\le K$ either contains $S$ (Step~2) or not (Step~3). Hence, for $N$ large,
\[
\Pp(\hat S_i\ne S_i)\le\Pp(G_N^c)+\Pp(\exists A:\ S\cup A\text{ beats }S,\ G_N)
\le3m^{-2}+C_5(\delta)\sum_{j\ge1}N^{j\{\kappa(1-\gamma)_+-1/2+C_6\delta\}},
\]
uniformly in $i$, which gives (a) since $m_N\to\infty$. For (b), $E_N$ is common to all $i$, and a union bound
over $i\le k_N\le m_N$ gives $\Pp(\bigcup_iG_N(i)^c)\le3m^{-1}\to0$. In Step~2 the union bound over nodes must
be taken before $m-1$ is replaced by $cN^\kappa$, because the positive part in the last display loses the
factor that offsets $k_N$ when $\gamma>1$. Put $M=m-1$ and $a=1-6\delta$. Since $k_N\le M+1$, the
contribution of sets $A$ with $j$ extra parents, summed over the nodes, is at most
\[
C\,M^{1+j(1-a\gamma)}N^{-aj/2}.
\]
If $1+j(1-a\gamma)\le0$ this is at most $CN^{-aj/2}\to0$. Otherwise $M\le cN^\kappa$ bounds it by
$C'N^{\kappa+j\{\kappa(1-a\gamma)-a/2\}}$. When $\gamma>2-1/(2\kappa)$ and $\delta$ is small enough,
$\kappa(1-a\gamma)-a/2<-\kappa<0$, so the exponent is largest at $j=1$, where it equals
$2\kappa-a(\kappa\gamma+1/2)<0$. The finitely many $j\le K$ therefore contribute $o(1)$, which gives (b).

\emph{Part (c).} Assumption (GA): with $t_j$ the $t$-statistic of $Z_j$, $j\in R_N$, in the regression of $Z_i$
on $Z_S$ and $Z_j$,
\[
\sup_u\Bigl|\Pp\bigl(\max_{j\in R_N}t_j\le u\bigr)-\Pp\bigl(\max_{j\in R_N}G_j\le u\bigr)\Bigr|\to0,
\]
where $G$ is a centred Gaussian vector with unit variances and the correlations of the population residuals of
the $Z_j$ given $Z_S$. Such approximations hold for maxima of high-dimensional sums under weak dependence
(Chernozhukov et al., 2013; Zhang and Wu, 2017); we assume it rather than verify their conditions. Fix $\eta>0$
with $2\kappa'(1-\bar\rho)>(1+\eta)(1+2\gamma\kappa)$ and let $u_N^2=(1+\eta)(1+2\gamma\kappa)\log N$, so that
$u_N^2\ge(1+\eta/2)\lambda_N$ for $N$ large. By Slepian's inequality (Slepian, 1962), $\Pp(\max_jG_j\le u_N)$ is
at most the same probability for an equicorrelated vector $H_j=\sqrt{\bar\rho}\,Z_0+\sqrt{1-\bar\rho}\,Z_j$
with independent standard normal $Z_0,Z_j$. Taking $a_N=\log\log N$,
\[
\Pp\bigl(\max_jH_j\le u_N\bigr)\le\Phi(-a_N)+\exp\{-|R_N|(1-\Phi(v_N))\},\qquad
v_N=\frac{u_N+\sqrt{\bar\rho}\,a_N}{\sqrt{1-\bar\rho}},
\]
and $|R_N|\{1-\Phi(v_N)\}\ge N^{\kappa'-(1+\eta)(1+2\gamma\kappa)/\{2(1-\bar\rho)\}+o(1)}\to\infty$. Hence
$\max_jt_j>u_N$ with probability tending to one. On that event let $j$ attain the maximum. Adding the lagged
edge $Z_j\to Z_i$ to $\Gs$ keeps the structure acyclic and changes the score by
$-N\log(1+t_j^2/\nu_j)+\lambda_N$, with $\nu_j=N-|S|-2$. Since $t_j>u_N>0$ and $x\mapsto\log(1+x)$ is
increasing,
\[
N\log(1+t_j^2/\nu_j)\ \ge\ N\log(1+u_N^2/\nu_j)=u_N^2\{1+o(1)\}\ \ge\ (1+\eta/2)\lambda_N\{1+o(1)\},
\]
where the $o(1)$ is deterministic because $u_N^2/N\to0$ and $\nu_j/N\to1$. So the change in score is negative
for $N$ large. All structures equivalent to $\Gs$ have the same score, so none of them minimises the score over
all structures. The added edge is lagged, so the new structure is compatible with every topological ordering
of $\Gs$, and the same conclusion holds over the structures compatible with a fixed such ordering. For the
size-capped estimator $\hat S_i$ of Proposition~3 the new parent set has $|S_i|+1$ elements, so the argument
needs $|S_i|<K$.
\end{proof}

%=====================================================================
\section{Index of simulation studies}\label{S-sec:index}

Table~\ref{S-tab:index} lists the studies behind the main paper. ``Full'' is the dynamic programme over all
orderings with backward elimination (Algorithm~1); ``local'' is the local search over orderings with the
stated number of random restarts added to the greedy start. Designs are drawn with \f{random_svar}: R7 uses
\f{parents = 1.5, off1 = 12, off2 = 9} and seeds 7001--7020, K15 uses seed 15015, and the larger designs the
default settings with seeds $1000k+d$. Series are generated after a burn-in of 500. Model~1 has
\[
A_0=\begin{pmatrix}1&0&0\\-1&1&0\\1&0&1\end{pmatrix},\quad
A_1=\begin{pmatrix}0.9&0&0\\0&0.6&0\\0&0&-0.5\end{pmatrix},
\]
\[
A_2=\begin{pmatrix}-0.6&0&0\\0.9&0&0.4\\0&0&0\end{pmatrix},
\]
in the sign convention of equation (1) of the paper, and
$D=\mathrm{diag}(0.09,0.04,0.025)$. Replication $r$ of R7 uses
design $((r-1)\bmod 20)+1$. The same cell can appear with different values in two studies because they use
different series, numbers of replications or searches: for R7 at $n=500$ with the BIC and the extended BIC,
8.6 and 8.8 (Table~1: 100 replications, full search), 8.8 and 8.5 (Table~2: the 20 exported series) and 7.4
and 8.0 (Table~3: 20 other series, local search without restarts).

{\footnotesize\setlength{\tabcolsep}{3.5pt}
\begin{longtable}{L{2.3cm}L{4.1cm}L{1.6cm}L{2.0cm}L{1.4cm}L{3.3cm}}
\caption{Simulation studies of the main paper.}\label{S-tab:index}\\
\toprule
Where reported & Designs, sample sizes and series & Replications & Search & Penalty & Script\\
\midrule
\endfirsthead
\toprule
Where reported & Designs, sample sizes and series & Replications & Search & Penalty & Script\\
\midrule
\endhead
\bottomrule
\endfoot
Table 1; Table~\ref{S-tab:smallfull} & Model 1, $n=100$ to 2000; R7, $n=200$ to 10\,000 & 200; 100 & full & BIC,
ext.\ BIC & \f{dag_sim.R}, \f{dag_sim_ebic.R}; R7 class recovery and split errors \f{r7_split_050.R}\\
\S6.4 ($k=15$) & K15, $n=300$ to 10\,000 & 60 & local, 3 restarts & BIC, ext.\ BIC & \f{dag_sim.R},
\f{dag_sim_ebic.R}\\
Table 2 & exported series: Model 1, $n=500$; R7, $n=500$, 2000 (one per design); K15, $n=1000$ & 20; 20; 20;
10 & ours: full ($k\le7$), local with 3 restarts ($k=15$) & BIC, ext.\ BIC & \f{export_tig*.R},
\f{ours_on_tig.R}, \f{run_tges.R}, \f{run_dynotears.py} and the scripts of each package\\
\S6.2 (TS-BOSS); Table~\ref{S-tab:tsboss} & R7 and dense lags (seeds 990000+, 991000+), $n=500$, 2000; $k=20$
series of Table~4 & 100; 100; 18 & ours: full ($k=7$), local with 3 restarts ($k=20$); TS-BOSS & BIC, ext.\ BIC &
\f{tsboss_compare.R}, \f{tsboss_run.py}, \f{tsboss_table.R}\\
\S6.2 (BiDAG) & R7 generator with one lag, seeds 7101--7120; $n=500$, 2000 & 20 & full & BIC, ext.\ BIC &
\f{run_bidag.R}\\
Table 3; Tables~\ref{S-tab:crosseddetail}--\ref{S-tab:k50} & series of Table~4: R7; $k=20$ (18 designs);
$k=30$ (6, $n=1000$); $k=50$ (12, $n=1000$) & one series per design and $n$ & local, no restarts; 3 restarts
at $k=50$; arc-wise searches & BIC, ext.\ BIC & \f{arcsearch.R}, \f{crossed.R}, \f{crossed_budget.R},
\f{k50_restarts.R}\\
Table 4 & $k=20$ (18 designs), 25 (12), 30 (6), 50 (12); $n=300$, 1000, 3000 ($k=50$: 1000); series seed
$77\,000+100d+\mathrm{index}(n)+10k$ & one series per design and $n$ & local, 3 restarts (version 0.5.0) & BIC, ext.\ BIC,
ANW screen & \f{table4_050.R} (series of \f{ebic_sim.R}, \f{ebic_more.R})\\
\S6.4 (BIC of the true DAG; single null lagged additions) & $k=30$, 6 designs, three sample sizes (other
series than Table~4) & 18 cells & local, 3 restarts (version 0.5.0) & BIC & \f{k30_check_050.R},
\f{k30_single_null.R}\\
\S5 (partial correlations, spectral radius) & R7 designs, population values & 20 designs & -- & -- &
\f{r7_partial_cor.R}\\
Table 5; Table~\ref{S-tab:gamma} & series of Table~1, of the $k=15$ study and of Table~4 & as those & as those &
$\gamma=0$ to 1 & \f{gamma_small.R}; \f{table5_gamma.R} for $k\ge20$ (version 0.5.0)\\
Table 6; Tables~\ref{S-tab:density}--\ref{S-tab:designs} & contemporaneous graphs of R7 with other lag
structures; $n=500$, 2000 & 100 (5 per design) & full; also on VAR residuals & BIC, ext.\ BIC &
\f{lag_density.R}, \f{lag_designs.R}\\
\S6.6; Table~\ref{S-tab:order} & Model 1, R7, K15 & 200; 40; 20 & software default & ext.\ BIC &
\f{order_sim.R} with $\gamma=0.5$\\
\S3.2, \S3.3, \S4.1; Section~\ref{S-sec:checks} & series of Table~1; random designs $k=8$ to 10; K15 & see
Section~\ref{S-sec:checks} & full and local & BIC, ext.\ BIC & \f{check_best_subset.R},
\f{check_exact_k10.R}, \f{check_full_k15.R}, \f{tests_heur*.R}\\
\S7; Section~\ref{S-sec:app} & six US series; series simulated from the fit, $n=202$ & 2000 resamples; 100
series & full & ext.\ BIC & \f{real_data_gmsvar.R}, \f{boot_calib050.R}, \f{macro_checks.R},
\f{macro_witness.R}\\
\end{longtable}}

Scripts and result files: \url{https://github.com/MarcoNZreal/gmsvar-replication}. Software: R 4.3.3; \texttt{gmsvar} 0.5.0;
pcalg 2.7-12; tpc 1.0; bnlearn 5.2.1; tigramite 5.2.10.1; causalnex 0.12.2; BiDAG 2.1.4 (with Rgraphviz 2.56.0);
TGES from \url{https://github.com/tobiaselar/TGES} (commit 2c40a66).

%=====================================================================
\section{Small systems: all restrictions of the candidates}\label{S-sec:small}

Table~\ref{S-tab:smallfull} extends Table~1 to the restrictions we tried. The graph of the window given to the
search is: the pairs whose partial correlation is significant at the 5\,\% level ($t$-test); the SIN rule, a
\v{S}id\'ak--Holm step-down procedure on Fisher $z$ statistics at family level 0.05
\citep{drton2007}; the graphical lasso with the penalty chosen by an extended BIC; an adaptive graphical lasso
with weights from the partial correlations (AGL); and an adaptive node-wise lasso (ANW). PC-ts is a tiered
PC-stable learner of our own, in the spirit of PCMCI$^{+}$, run on the same series; it is not a restriction
of the search but a learner in its own right.

\begin{table}[h]\centering\footnotesize\setlength{\tabcolsep}{4pt}
\begin{tabular}{lrrrrrrrrrrr}
\toprule
 & $n$ & None & None & $t$-test & SIN & glasso & AGL & ANW & Oracle & PC-ts & PC-ts\\
 & & BIC & ext.\ BIC & & & & & & & .01 & .05\\
\midrule
\multicolumn{12}{l}{\emph{Model 1: share of runs recovering the true class}}\\
 & 100 & 0.52 & 0.74 & 0.14 & 0.01 & 0.14 & 0.45 & 0.33 & 0.89 & 0.00 & 0.01\\
 & 200 & 0.76 & 0.91 & 0.44 & 0.12 & 0.67 & 0.74 & 0.57 & 0.95 & 0.01 & 0.18\\
 & 500 & 0.84 & 0.94 & 0.77 & 0.45 & 0.84 & 0.84 & 0.81 & 0.95 & 0.46 & 0.67\\
 & 2000 & 0.96 & 0.99 & 0.99 & 0.99 & 0.96 & 0.98 & 0.99 & 0.99 & 0.96 & 0.78\\
\multicolumn{12}{l}{\emph{R7: mean skeleton errors among 168 cells}}\\
 & 200 & 19.4 & 19.3 & 23.7 & 33.1 & 42.7 & 21.4 & 20.8 & 11.6 & 28.5 & 25.4\\
 & 500 & 8.6 & 8.8 & 13.6 & 23.7 & 19.9 & 12.4 & 11.7 & 5.0 & 19.1 & 16.8\\
 & 2000 & 1.3 & 0.9 & 5.5 & 9.7 & 5.1 & 5.6 & 5.4 & 0.6 & 8.7 & 8.5\\
 & 10\,000 & 0.4 & 0.1 & 3.2 & 4.7 & 1.6 & 4.3 & 4.7 & 0.1 & 4.2 & 5.2\\
\bottomrule
\end{tabular}
\caption{Full search without a restriction (BIC and extended BIC) and, with the BIC, restricted to an
estimated graph of the window or to the true moral graph (oracle); PC-ts at levels 0.01 and 0.05. 200
replications for Model~1 and 100 for R7.}\label{S-tab:smallfull}
\end{table}

The order is the same at every $n$ in R7: oracle, no restriction, then the estimated graphs, with the most
conservative rule (SIN) the worst, because conservative rules drop true edges (lag-to-current true-positive
rate at $n=2000$: 0.94 for the $t$-test, 0.83 for SIN, 0.99 unrestricted). The penalised graphs are no better
than none. In Model~1 at $n=2000$ the restricted searches are slightly better than the unrestricted BIC
search (0.99 against 0.96), because a graph that retains the true edges removes spurious candidates; the
extended BIC achieves the same without a screen.

For the $k=15$ design (local search, three restarts, 60 replications) the unrestricted search makes 21.0, 7.2,
2.9 and 1.5 skeleton errors among 555 cells with the BIC at $n=300$, 1000, 3000 and 10\,000, and 12.7, 3.4,
0.65 and 0.13 with the extended BIC, which recovers the class in 0.45 and 0.88 of the runs at $n=3000$ and
10\,000 (0.07 and 0.30 with the BIC). With the true moral graph the BIC makes 6.8, 1.8, 0.3 and 0.15 errors.

%=====================================================================
\section{Sensitivity to the extended-BIC parameter}\label{S-sec:gamma}

\begin{table}[h]\centering\small
\begin{tabular}{llrrrrrr}
\toprule
Design & measure & $n$ & $\gamma=0$ (BIC) & 0.25 & 0.5 & 0.75 & 1\\
\midrule
Model 1 & class & 100 & 0.52 & 0.65 & 0.74 & 0.79 & 0.81\\
 & & 500 & 0.84 & 0.91 & 0.94 & 0.96 & 0.97\\
 & & 2000 & 0.96 & 0.99 & 0.99 & 0.99 & 0.99\\
R7 & errors & 200 & 19.4 & 18.4 & 19.3 & 20.2 & 21.3\\
 & & 500 & 8.6 & 8.4 & 8.8 & 9.4 & 10.2\\
 & & 2000 & 1.3 & 1.0 & 0.9 & 1.1 & 1.3\\
 & & 10\,000 & 0.4 & 0.1 & 0.1 & 0.0 & 0.0\\
$k=15$ & errors & 300 & 21.0 & 14.1 & 12.7 & 12.7 & 13.2\\
 & & 1000 & 7.2 & 4.2 & 3.4 & 3.5 & 3.8\\
 & & 3000 & 2.9 & 1.2 & 0.7 & 0.5 & 0.5\\
 & & 10\,000 & 1.5 & 0.5 & 0.1 & 0.1 & 0.0\\
$k=20$ & errors & 300 & 36.3 & 24.8 & 22.3 & 22.7 & 22.8\\
 & & 1000 & 13.2 & 6.9 & 4.7 & 4.3 & 4.5\\
 & & 3000 & 5.7 & 1.9 & 1.0 & 0.4 & 0.4\\
$k=25$ & errors & 300 & 57.3 & 33.1 & 27.3 & 27.5 & 30.7\\
 & & 1000 & 19.8 & 8.6 & 6.6 & 6.2 & 6.3\\
 & & 3000 & 7.5 & 2.5 & 1.0 & 0.4 & 0.5\\
$k=30$ & errors & 300 & 78.7 & 48.3 & 37.8 & 35.5 & 37.5\\
 & & 1000 & 26.8 & 9.3 & 6.3 & 6.5 & 6.7\\
 & & 3000 & 12.0 & 5.2 & 1.5 & 0.8 & 0.7\\
$k=50$ & errors & 1000 & 69.7 & 20.8 & 11.7 & 10.7 & 12.5\\
\bottomrule
\end{tabular}
\caption{Sensitivity to $\gamma$: class recovery (Model~1) or mean skeleton errors (other designs) of the
unrestricted search. Model~1: 200 replications; R7: 100; $k=15$: 60; $k=20$, 25, 30, 50: 18, 12, 6 and 12
designs with one series each, local search with three restarts (the series and search of Table~4).}\label{S-tab:gamma}
\end{table}

Table~\ref{S-tab:gamma} is the full version of Table~5 of the paper. Every $\gamma>0$ improves on the BIC from
$k=15$ on. $\gamma=0.25$ removes most of the
excess errors of the BIC but not all (at $k=50$: 20.8 errors against 69.7 for the BIC and 10.7 to 12.5 for
$\gamma\ge0.5$). Between 0.5 and 1 the results are flat in the larger systems, within three and a half errors. Larger $\gamma$ helps in Model~1 but costs in R7 at small $n$, where many true
links are weak. No value is best everywhere; 0.5 is never far from the best and keeps the weak-signal designs
close to the BIC.

%=====================================================================
\section{Implementation checks}\label{S-sec:checks}

\emph{Backward elimination against the best subset.} On series from the designs of Table~1 (200 per sample
size in Model~1, 40 in R7) the full search was rerun with each node problem solved exactly by branch and
bound (package \texttt{leaps}), with both penalties. Backward elimination returned the best subset in 94 to
100\,\% of the node problems in Model~1 and in 63 to 95\,\% in R7, and the final graph was the same in 89 to
100\,\% of the Model~1 fits and in 88 to 100\,\% of the R7 fits at $n\ge2000$, but in only 18\,\% of the R7
fits at $n=200$ and 38 to 58\,\% at $n=500$, where many true links are weak. The effect on accuracy is small:
in Model~1 the best subset lowers the skeleton errors from 1.10 to 0.85 at $n=100$ with the BIC (0.62 to 0.51
with the extended BIC); in R7 the two are within about half an error of each other at every $n$, in either
direction at $n\le500$. With the extended BIC at the largest sample sizes the two agree in every fit. This is
the finite-sample counterpart of Proposition~2: the full search is not an exact minimiser of the score, and
is consistent nevertheless.

\emph{Local search against the full search.} The local search with three restarts returned the parent sets
of the full search in 40 of 40 fits in Model~1 and in 118 of 120 in R7 (the two exceptions, at $n=200$, with
a BIC at most 5.3 units worse); with the extended BIC in 59 of 60 fits on random designs with $k=8$, 9 and 10
(the exception 1.1 units worse), and in 12 of 12 fits on K15 series ($n=300$ and 1000, both penalties), where
the full search takes three to four minutes. At $k=50$ restarts matter: see Section~\ref{S-sec:crossed}.

\emph{Other checks.} The $O(q^3)$ backward-elimination routine returns the parent sets of a reference routine
that re-solves each regression in all 2880 fits compared. The orientation rules agree with brute-force
enumeration of the equivalence class on 300 of 300 random window graphs. The lag-order criteria agree with
\texttt{vars::VARselect}.

%=====================================================================
\section{Published learners: settings and further results}\label{S-sec:learners}

All learners were given the true lag order and the series of Table~2. Skeleton errors are counted over the
pairs that involve a current variable; edges that a learner places among lagged variables are not scored.

\emph{PCMCI$^{+}$} (tigramite): partial-correlation tests, $\tau_{\max}=p$, level 0.01. \emph{Tiered PC}
(tpc): Gaussian tests at level 0.01, lags as tiers, conditioning sets of size at most three; it does not
force adjacency among lagged variables. \emph{PC and GES} (pcalg): applied to the window without the time
order. \emph{Hill climbing and tabu search} (bnlearn): BIC, arcs from newer to older lags blacklisted. At
$k=15$, tpc, pcalg and bnlearn were run on the first five of the ten series.

\emph{TGES.} Authors' code, Gaussian BIC score, window columns as variables with lag $l$ in tier $p+1-l$. It
models the joint distribution of the lagged variables as well as the current ones. It was worse than our search with the extended BIC in
90\,\% of the R7 series at $n=500$, 95\,\% at $n=2000$ and 90\,\% at $k=15$, and tied in 70\,\% of the
Model~1 series. In R7 its misses fall from 8.2 to 3.4 between $n=500$ and 2000 while its false inclusions rise
from 9.8 to 14.4; nearly all its errors are lagged links. A fit took 0.6\,s in Model~1, about 30\,s in R7 and
about 150\,s at $k=15$, the score being evaluated in R.

\emph{DYNOTEARS.} Implementation of causalnex, loaded without modification; equal penalties on the
contemporaneous and the lagged weights, of 0.01, 0.03 or 0.1, on centred series and on standardised series;
weights below a threshold of 0, 0.01, 0.03, 0.05 or 0.1 set to zero (Table~\ref{S-tab:dyn}). Without a threshold many negligible
weights are returned as edges. Even with the best setting per cell DYNOTEARS made 1.4 to 14.9 errors
against 0.2 to 8.5 for our search with the extended BIC (Table~2). On the simulated scale it gave the true
direction to 81 to 100\,\% of the contemporaneous edges, but to about half of them in R7 when the series were
standardised, whereas our search oriented 96 to 100\,\% correctly (Table~\ref{S-tab:direction}).

\begin{table}[h]\centering\small
\begin{tabular}{rrrrrr}
\toprule
Penalty & Threshold & Model 1, $n=500$ & R7, $n=500$ & R7, $n=2000$ & $k=15$, $n=1000$\\
\midrule
\multicolumn{6}{l}{\emph{raw series}}\\
0.01 & 0 & 4.2 & 102.6 & 84.8 & 387.4\\
0.01 & 0.01 & 3.4 & 81.9 & 52.5 & 271.2\\
0.01 & 0.03 & 2.5 & 54.0 & 24.4 & 138.3\\
0.01 & 0.05 & 1.9 & 36.6 & 15.4 & 73.6\\
0.01 & 0.1 & 1.4 & 19.1 & 13.1 & 39.2\\
0.03 & 0 & 6.8 & 70.7 & 48.0 & 242.1\\
0.03 & 0.01 & 6.2 & 53.1 & 27.9 & 144.2\\
0.03 & 0.03 & 5.8 & 34.0 & 13.7 & 68.7\\
0.03 & 0.05 & 5.5 & 21.6 & 10.9 & 43.1\\
0.03 & 0.1 & 4.5 & 16.8 & 11.9 & 31.7\\
0.1 & 0 & 6.2 & 32.5 & 25.6 & 79.3\\
0.1 & 0.01 & 6.1 & 23.6 & 15.6 & 39.4\\
0.1 & 0.03 & 5.8 & 17.3 & 11.9 & 18.9\\
0.1 & 0.05 & 5.5 & 14.6 & 11.8 & 14.9\\
0.1 & 0.1 & 5.3 & 16.6 & 14.8 & 17.4\\
\multicolumn{6}{l}{\emph{standardised series}}\\
0.01 & 0 & 7.8 & 79.0 & 63.9 & 211.7\\
0.01 & 0.01 & 6.3 & 62.2 & 45.0 & 128.7\\
0.01 & 0.03 & 4.8 & 44.1 & 28.8 & 67.5\\
0.01 & 0.05 & 4.4 & 31.9 & 22.0 & 43.7\\
0.01 & 0.1 & 4.3 & 22.1 & 20.2 & 30.5\\
0.03 & 0 & 8.0 & 46.7 & 36.0 & 98.4\\
0.03 & 0.01 & 6.4 & 36.0 & 26.8 & 60.9\\
0.03 & 0.03 & 6.0 & 27.0 & 20.2 & 38.6\\
0.03 & 0.05 & 5.8 & 22.2 & 17.1 & 30.4\\
0.03 & 0.1 & 4.8 & 21.6 & 19.8 & 27.4\\
0.1 & 0 & 5.6 & 28.2 & 24.6 & 41.9\\
0.1 & 0.01 & 4.7 & 25.4 & 22.1 & 32.5\\
0.1 & 0.03 & 4.5 & 24.6 & 21.9 & 25.8\\
0.1 & 0.05 & 4.5 & 24.1 & 22.9 & 24.2\\
0.1 & 0.1 & 3.9 & 26.4 & 25.6 & 29.6\\
\bottomrule
\end{tabular}
\caption{DYNOTEARS: mean skeleton errors for each penalty and weight threshold (generated).}\label{S-tab:dyn}
\end{table}

\begin{table}[h]\centering\small
\begin{tabular}{lrrrr}
\toprule
 & Model 1 & R7, $n=500$ & R7, $n=2000$ & $k=15$\\
\midrule
Our search, extended BIC & 100 / 0 & 96 / 4 & 100 / 0 & 100 / 0\\
TGES & 90 / 10 & 59 / 37 & 62 / 35 & 91 / 9\\
DYNOTEARS, penalty 0.1, threshold 0.05, centred series & 72 / 28 & 85 / 14 & 85 / 14 & 92 / 7\\
DYNOTEARS, best setting per cell, standardised series & 60 / 40 & 47 / 46 & 53 / 41 & 73 / 17\\
\bottomrule
\end{tabular}
\caption{Direction of the true contemporaneous edges, all of which are compelled in these designs: percentage
given the right and the wrong direction. For our search, the percentage compelled in each
direction.}\label{S-tab:direction}
\end{table}

\emph{BiDAG.} Its dynamic Bayesian networks are first order, so it was compared on a separate design: the R7
generator with one lag (20 designs, one series each). The data are one row per transition $(X_{t-1},X_t)$
with two slices; the score is BGe and the estimate is the maximum-scoring DAG of \f{iterativeMCMC()} with
default settings. With \f{samestruct = FALSE} the transition structure is learned on its own, which is the
model that matches a stationary series; the package default, \f{samestruct = TRUE}, assumes that the first
slice has the internal structure of the later ones, which does not hold for the margin of $X_{t-1}$
(Table~\ref{S-tab:bidag}). With \f{samestruct = FALSE} the paired differences from our search with the extended
BIC were 1.5 and 0.5 errors at $n=500$ and 2000 (standard errors 0.7 and 0.4), with similar orientation
accuracy, and a fit took about 8\,s against 0.04\,s.

\begin{table}[h]\centering\footnotesize\setlength{\tabcolsep}{4pt}
\begin{tabular}{lrrrrrrr}
\toprule
 & \multicolumn{2}{c}{Skeleton errors} & \multicolumn{2}{c}{Contemporaneous errors} &
\multicolumn{2}{c}{Arrows right / wrong (\%)} & Seconds\\
 & $n=500$ & 2000 & $n=500$ & 2000 & $n=500$ & 2000 & per fit\\
\midrule
Our search, extended BIC & 2.3 & 0.5 & 0.3 & 0.0 & 95 / 4 & 99 / 1 & 0.04\\
Our search, BIC & 2.0 & 0.8 & 0.4 & 0.2 & 97 / 2 & 99 / 1 & 0.04\\
BiDAG, transition structure only & 3.7 & 0.9 & 0.5 & 0.1 & 89 / 8 & 98 / 2 & 8--9\\
BiDAG, same structure (default) & 8.2 & 7.6 & 3.7 & 5.8 & 86 / 11 & 93 / 5 & 7--8\\
\bottomrule
\end{tabular}
\caption{BiDAG and our search on twenty seven-variable designs with one lag. BiDAG (transition only) was worse
than our search with the extended BIC in 45\,\% of the designs at $n=500$ (tie 50\,\%, better 5\,\%) and in
20\,\% at $n=2000$ (tie 75\,\%, better 5\,\%).}\label{S-tab:bidag}
\end{table}

\emph{TS-BOSS.} The code of \citet{castillo2026} (\url{https://github.com/irenecastillo/TS-BOSS}, commit
d0cf643, MIT licence) was run in Python 3.12 on the window matrix used by our search, with its columns in the
order of that code (oldest lag first), so that both use the same $N=n-p$ rows. Its score is
$-\log L_{xx}-|Z|\,\mathrm{pd}\log N/(4N)$ per node, with $L$ the Cholesky factor of the correlation matrix,
which equals $-1/(2N)$ times our $s_i(P)$ with $\lambda_N=(\mathrm{pd}/2)\log N$ up to a constant; we set
$\mathrm{pd}=2$ (BIC) and $\mathrm{pd}=2\lambda_N/\log N$ with $\lambda_N$ of equation (3) of the paper. (The
constructor converts pd to an integer, so we set it on the object.) The released code runs the search over
orderings with grow--shrink trees and returns its DAG; it has no backward equivalence search. We computed the
class of every DAG, ours and TS-BOSS's, with the same rules. Our search was the full one for $k=7$ and the
local one with three restarts for $k=20$. Dense lags: the contemporaneous graphs of R7 with one lag and 28 of
42 cross-lag cells non-zero (Table~6), fitted with $p=3$; five series per design. The $k=20$ series are those
of Table~4. Table~\ref{S-tab:tsboss} gives the results; the standard error of the paired difference is computed
across designs: for R7 and dense lags from the 20 design means of the five paired differences, for $k=20$
from the 18 designs.

\begin{table}[h]\centering\footnotesize\setlength{\tabcolsep}{3.6pt}
\begin{tabular}{llrrrrrrrr}
\toprule
 & & & \multicolumn{3}{c}{Skeleton errors} & Precision & Recall & Lower score & Seconds\\
Penalty & Design & $n$ & ours & TS-BOSS & diff.\ (s.e.) & ours / TS & ours / TS & ours / TS & ours / TS\\
\midrule
BIC & R7 & 500 & 8.2 & 8.9 & -0.70 (0.34) & 93 / 89 & 96 / 93 & 32 / 15 & 0.11 / 0.39\\
BIC & R7 & 2000 & 1.3 & 1.4 & -0.12 (0.07) & 98 / 98 & 100 / 100 & 4 / 0 & 0.10 / 0.44\\
BIC & Dense lags & 500 & 12.5 & 12.3 & 0.20 (0.44) & 54 / 55 & 60 / 62 & 44 / 10 & 0.11 / 0.43\\
BIC & Dense lags & 2000 & 3.3 & 3.9 & -0.57 (0.51) & 87 / 81 & 89 / 85 & 25 / 2 & 0.11 / 0.52\\
BIC & $k=20$ & 300 & 36.3 & 37.6 & -1.28 (2.48) & 85 / 78 & 98 / 94 & 56 / 44 & 9.50 / 17.89\\
BIC & $k=20$ & 1000 & 13.2 & 12.5 & 0.67 (0.52) & 91 / 92 & 99 / 99 & 39 / 33 & 10.41 / 31.07\\
BIC & $k=20$ & 3000 & 5.7 & 5.9 & -0.22 (0.60) & 95 / 94 & 100 / 100 & 17 / 11 & 11.58 / 39.19\\
\addlinespace
ext.\ BIC & R7 & 500 & 8.4 & 9.8 & -1.33 (0.55) & 94 / 89 & 95 / 91 & 49 / 15 & 0.11 / 0.33\\
ext.\ BIC & R7 & 2000 & 0.9 & 1.0 & -0.11 (0.09) & 100 / 99 & 100 / 100 & 3 / 0 & 0.10 / 0.43\\
ext.\ BIC & Dense lags & 500 & 12.2 & 11.8 & 0.43 (0.48) & 51 / 56 & 56 / 60 & 54 / 6 & 0.11 / 0.41\\
ext.\ BIC & Dense lags & 2000 & 3.0 & 3.9 & -0.82 (0.52) & 86 / 79 & 87 / 82 & 24 / 0 & 0.11 / 0.49\\
ext.\ BIC & $k=20$ & 300 & 22.3 & 23.8 & -1.56 (1.60) & 94 / 92 & 96 / 95 & 83 / 17 & 8.81 / 7.21\\
ext.\ BIC & $k=20$ & 1000 & 4.7 & 5.3 & -0.61 (0.71) & 98 / 98 & 100 / 100 & 44 / 17 & 9.49 / 20.98\\
ext.\ BIC & $k=20$ & 3000 & 1.0 & 1.5 & -0.50 (0.44) & 99 / 99 & 100 / 100 & 11 / 0 & 11.38 / 33.77\\
\bottomrule
\end{tabular}
\caption{TS-BOSS and our search on identical window matrices with the same penalty. Precision: percentage of
the contemporaneous arrows reported as compelled that are compelled in that direction in truth; recall:
percentage of the truly compelled arrows compelled in the right direction; lower score: percentage of series
in which each search reached the strictly lower score (the rest tie). Diff.: mean of ours minus TS-BOSS,
with its standard error across designs. R7 and dense lags: 100 series per cell;
$k=20$: 18. Times are for R (ours) and Python (TS-BOSS) on one core.}\label{S-tab:tsboss}
\end{table}

%=====================================================================
\section{Search and penalty with a common score}\label{S-sec:crossed}

The arc-wise searches of Table~3 minimise the score (2) over the same structures as our search: lagged
vertices have no parents, lag-to-current arcs are always permitted and the contemporaneous arcs must be
acyclic. \emph{Hill climbing} starts from the empty graph and applies the best addition, deletion or reversal
of one arc until none lowers the score; it was checked by brute force to end at a local optimum.
\emph{Tabu search} uses the same moves with a tabu list of ten structures and stops after ten steps without
improvement. \emph{Restarted hill climbing} restarts from random perturbations of the best graph, of
$\max(3,k/6)$ random arc changes, until the elapsed time of the search over orderings on the same series is
used up: for $k\le10$ the larger of the times of the local and the full search (in R7 the full search, about
0.10\,s), otherwise that of the local search. The restart in progress is completed, so the elapsed time can
exceed the budget by one restart. A
further variant starts hill climbing with all lagged variables as parents. Table~\ref{S-tab:crosseddetail} gives
errors, false inclusions and misses, the score relative to that of the true graph and the time.

{\footnotesize
\begin{longtable}{llllrrrrr}
\caption{Common-score comparison in detail: means over designs (generated). ``Score'' is the score of the
estimate minus that of the true graph.}\label{S-tab:crosseddetail}\\
\toprule
Design & $n$ & Penalty & Search & Errors & False incl. & Misses & Score & Seconds\\
\midrule
\endfirsthead
\toprule
Design & $n$ & Penalty & Search & Errors & False incl. & Misses & Score & Seconds\\
\midrule
\endhead
\bottomrule
\endfoot
R7 & 500 & BIC & ordering, local & 7.4 & 2.9 & 4.5 & -15 & 0.05\\
R7 & 500 & BIC & arc, hill climbing & 18.6 & 9.3 & 9.3 & 25 & 0.01\\
R7 & 500 & BIC & arc, tabu & 18.7 & 9.7 & 9.0 & 22 & 0.05\\
R7 & 500 & BIC & arc, restarted, equal time & 17.3 & 8.6 & 8.7 & 15 & 0.10\\
R7 & 500 & BIC & arc, all lags at start & 19.8 & 12.5 & 7.3 & 30 & 0.03\\
R7 & 500 & ext.\ BIC & ordering, local & 8.0 & 1.6 & 6.5 & -28 & 0.04\\
R7 & 500 & ext.\ BIC & arc, hill climbing & 17.8 & 5.8 & 12.0 & 20 & 0.01\\
R7 & 500 & ext.\ BIC & arc, tabu & 16.2 & 5.4 & 10.9 & 7 & 0.05\\
R7 & 500 & ext.\ BIC & arc, restarted, equal time & 15.8 & 5.2 & 10.6 & 3 & 0.10\\
R7 & 500 & ext.\ BIC & arc, all lags at start & 18.7 & 8.8 & 9.9 & 31 & 0.03\\
\addlinespace
R7 & 2000 & BIC & ordering, local & 1.8 & 1.3 & 0.5 & -3 & 0.04\\
R7 & 2000 & BIC & arc, hill climbing & 12.2 & 9.4 & 2.9 & 55 & 0.02\\
R7 & 2000 & BIC & arc, tabu & 11.1 & 8.7 & 2.4 & 48 & 0.06\\
R7 & 2000 & BIC & arc, restarted, equal time & 10.6 & 8.1 & 2.6 & 41 & 0.10\\
R7 & 2000 & BIC & arc, all lags at start & 17.9 & 15.4 & 2.6 & 90 & 0.03\\
R7 & 2000 & ext.\ BIC & ordering, local & 1.3 & 0.4 & 0.9 & -3 & 0.04\\
R7 & 2000 & ext.\ BIC & arc, hill climbing & 10.2 & 6.9 & 3.4 & 76 & 0.02\\
R7 & 2000 & ext.\ BIC & arc, tabu & 9.2 & 6.2 & 3.0 & 63 & 0.06\\
R7 & 2000 & ext.\ BIC & arc, restarted, equal time & 8.7 & 5.7 & 3.0 & 60 & 0.10\\
R7 & 2000 & ext.\ BIC & arc, all lags at start & 14.8 & 11.4 & 3.4 & 111 & 0.03\\
\addlinespace
$k=$20 & 300 & BIC & ordering, local & 40.4 & 28.5 & 11.9 & -38 & 4.45\\
$k=$20 & 300 & BIC & arc, hill climbing & 54.8 & 35.6 & 19.2 & 51 & 0.17\\
$k=$20 & 300 & BIC & arc, tabu & 55.6 & 36.6 & 19.0 & 49 & 0.56\\
$k=$20 & 300 & BIC & arc, restarted, equal time & 53.6 & 36.3 & 17.3 & 24 & 4.46\\
$k=$20 & 300 & BIC & arc, all lags at start & 72.6 & 55.9 & 16.7 & 67 & 1.20\\
$k=$20 & 300 & ext.\ BIC & ordering, local & 23.5 & 6.4 & 17.1 & -61 & 4.11\\
$k=$20 & 300 & ext.\ BIC & arc, hill climbing & 42.5 & 16.4 & 26.1 & 71 & 0.13\\
$k=$20 & 300 & ext.\ BIC & arc, tabu & 41.8 & 16.6 & 25.2 & 65 & 0.49\\
$k=$20 & 300 & ext.\ BIC & arc, restarted, equal time & 37.7 & 14.6 & 23.1 & 27 & 4.12\\
$k=$20 & 300 & ext.\ BIC & arc, all lags at start & 44.2 & 22.0 & 22.2 & 79 & 1.18\\
\addlinespace
$k=$20 & 1000 & BIC & ordering, local & 13.6 & 11.6 & 2.0 & -11 & 4.69\\
$k=$20 & 1000 & BIC & arc, hill climbing & 40.7 & 34.7 & 6.0 & 174 & 0.22\\
$k=$20 & 1000 & BIC & arc, tabu & 40.7 & 35.0 & 5.7 & 171 & 0.76\\
$k=$20 & 1000 & BIC & arc, restarted, equal time & 33.6 & 28.7 & 4.8 & 117 & 4.70\\
$k=$20 & 1000 & BIC & arc, all lags at start & 58.6 & 53.8 & 4.7 & 243 & 1.34\\
$k=$20 & 1000 & ext.\ BIC & ordering, local & 6.1 & 3.2 & 2.9 & 3 & 4.47\\
$k=$20 & 1000 & ext.\ BIC & arc, hill climbing & 31.7 & 23.1 & 8.6 & 269 & 0.21\\
$k=$20 & 1000 & ext.\ BIC & arc, tabu & 31.1 & 22.8 & 8.3 & 260 & 0.70\\
$k=$20 & 1000 & ext.\ BIC & arc, restarted, equal time & 25.4 & 18.6 & 6.8 & 197 & 4.47\\
$k=$20 & 1000 & ext.\ BIC & arc, all lags at start & 35.8 & 28.6 & 7.2 & 293 & 1.40\\
\addlinespace
$k=$20 & 3000 & BIC & ordering, local & 6.8 & 6.4 & 0.3 & 0 & 4.45\\
$k=$20 & 3000 & BIC & arc, hill climbing & 38.4 & 36.3 & 2.2 & 258 & 0.26\\
$k=$20 & 3000 & BIC & arc, tabu & 37.8 & 35.7 & 2.1 & 249 & 0.89\\
$k=$20 & 3000 & BIC & arc, restarted, equal time & 34.9 & 33.1 & 1.9 & 220 & 4.46\\
$k=$20 & 3000 & BIC & arc, all lags at start & 59.2 & 57.2 & 1.9 & 393 & 1.35\\
$k=$20 & 3000 & ext.\ BIC & ordering, local & 1.6 & 1.2 & 0.3 & 5 & 4.46\\
$k=$20 & 3000 & ext.\ BIC & arc, hill climbing & 30.7 & 27.8 & 2.8 & 377 & 0.23\\
$k=$20 & 3000 & ext.\ BIC & arc, tabu & 28.7 & 26.1 & 2.6 & 353 & 0.81\\
$k=$20 & 3000 & ext.\ BIC & arc, restarted, equal time & 28.8 & 26.4 & 2.4 & 345 & 4.47\\
$k=$20 & 3000 & ext.\ BIC & arc, all lags at start & 46.4 & 44.1 & 2.4 & 558 & 1.38\\
\addlinespace
$k=$30 & 1000 & BIC & ordering, local & 26.8 & 23.7 & 3.2 & -24 & 19.72\\
$k=$30 & 1000 & BIC & arc, hill climbing & 54.7 & 47.2 & 7.5 & 203 & 0.51\\
$k=$30 & 1000 & BIC & arc, tabu & 55.7 & 48.2 & 7.5 & 203 & 1.43\\
$k=$30 & 1000 & BIC & arc, restarted, equal time & 47.7 & 41.7 & 6.0 & 127 & 19.75\\
$k=$30 & 1000 & BIC & arc, all lags at start & 94.0 & 87.2 & 6.8 & 378 & 4.99\\
$k=$30 & 1000 & ext.\ BIC & ordering, local & 8.0 & 2.8 & 5.2 & -9 & 19.43\\
$k=$30 & 1000 & ext.\ BIC & arc, hill climbing & 38.5 & 27.5 & 11.0 & 331 & 0.44\\
$k=$30 & 1000 & ext.\ BIC & arc, tabu & 38.7 & 27.8 & 10.8 & 328 & 1.34\\
$k=$30 & 1000 & ext.\ BIC & arc, restarted, equal time & 21.0 & 14.0 & 7.0 & 117 & 19.44\\
$k=$30 & 1000 & ext.\ BIC & arc, all lags at start & 48.2 & 37.8 & 10.3 & 432 & 5.15\\
\addlinespace
\end{longtable}}

\emph{More time for the arc-wise search.} In R7, restarted hill climbing with 10 and 100 times the time of the
full search made 14.3 and 9.9 errors with the BIC at $n=500$ (full search 7.4), 8.0 and 3.7 at $n=2000$ (1.8);
with the extended BIC 13.7 and 11.4 at $n=500$ (7.7) and 5.8 and 1.7 at $n=2000$ (1.3). With 100 times the
time it ended 2.6 to 8.5 score units above the full search on average.

\emph{Fifty variables.} Table~\ref{S-tab:k50} compares, design by design, the local search over orderings with
three restarts and restarted hill climbing given the same time. The paired difference is 2.8 errors (standard
error 2.3). The search over orderings has fewer errors in four designs, the arc-wise search in five, with
three ties; the arc-wise search reaches the lower score in eight designs. The mean favours the search over
orderings because of designs 1 and 5, where the arc-wise search ends far above the true graph's score.
These are the series of Table~4, whose extended-BIC entry at $k=50$ is the same 11.7; without restarts the
search over orderings made 13.2 errors.

\begin{table}[h]\centering\small
\begin{tabular}{rrrrrr}
\toprule
 & \multicolumn{2}{c}{Search over orderings, 3 restarts} & \multicolumn{2}{c}{Arc-wise, equal time} & \\
Design & Errors & Score & Errors & Score & Seconds\\
\midrule
1 & 6 & -25.5 & 26 & 150.6 & 322\\
2 & 15 & -48.4 & 21 & -17.3 & 293\\
3 & 6 & -10.8 & 6 & -10.8 & 400\\
4 & 24 & -84.2 & 23 & -92.7 & 443\\
5 & 7 & -15.5 & 24 & 69.5 & 471\\
6 & 7 & -23.8 & 7 & -25.3 & 540\\
7 & 12 & -43.9 & 10 & -50.3 & 530\\
8 & 23 & -30.3 & 17 & -46.9 & 573\\
9 & 11 & -38.9 & 9 & -48.2 & 500\\
10 & 8 & -24.2 & 8 & -24.8 & 509\\
11 & 13 & -23.5 & 12 & -26.3 & 335\\
12 & 8 & -31.2 & 10 & -31.9 & 300\\
\midrule Mean & 11.7 & -33.4 & 14.4 & -12.9 & 435\\
\bottomrule
\end{tabular}
\caption{Fifty variables, $n=1000$, extended BIC (generated). ``Score'' is the score of the estimate minus
that of the true graph.}\label{S-tab:k50}
\end{table}

%=====================================================================
\section{Lag structure and orientation}\label{S-sec:lags}

\emph{Lag density.} Table~\ref{S-tab:density} gives, for the designs of Table~6, the orientation of the truly
compelled arrows under five procedures: the unrestricted search with the extended BIC and with the BIC; the
search restricted by a $t$-test screen; the unrestricted extended-BIC fit with its arrows read from the
contemporaneous graph only, ignoring its lagged links; and the same search applied to the residuals of the
full VAR(3), which uses no lag exclusion. The last column is the share of the truly free edges, where there
are any, that the unrestricted fit reports as compelled.

\begin{table}[h]\centering\footnotesize\setlength{\tabcolsep}{3.2pt}
\begin{tabular}{rrrrrrrrrrrr}
\toprule
Cells & \multicolumn{2}{c}{Ext.\ BIC} & \multicolumn{2}{c}{BIC} & \multicolumn{2}{c}{$t$-test screen} &
\multicolumn{2}{c}{Contemp.\ only} & \multicolumn{2}{c}{VAR residuals} & Free edges\\
(of 42) & right & wrong & right & wrong & right & wrong & right & wrong & right & wrong & compelled\\
\midrule
\multicolumn{12}{l}{\emph{$n=500$}}\\
0 & 100 & 0 & 100 & 0 & 93 & 4 & 28 & 0 & 43 & 15 & --\\
7 & 97 & 2 & 97 & 2 & 86 & 11 & 29 & 0 & 43 & 14 & --\\
14 & 93 & 6 & 93 & 6 & 78 & 18 & 30 & 2 & 41 & 14 & --\\
21 & 78 & 18 & 81 & 16 & 67 & 26 & 32 & 7 & 40 & 16 & 80\\
28 & 54 & 41 & 61 & 35 & 56 & 39 & 25 & 17 & 42 & 16 & 100\\
35 & 50 & 45 & 54 & 43 & 50 & 45 & 24 & 16 & 42 & 14 & 91\\
42 & 35 & 65 & 41 & 59 & 48 & 52 & 16 & 20 & 91 & 4 & 91\\
\multicolumn{12}{l}{\emph{$n=2000$}}\\
0 & 100 & 0 & 100 & 0 & 96 & 2 & 28 & 0 & 37 & 11 & --\\
7 & 100 & 0 & 100 & 0 & 92 & 6 & 28 & 0 & 37 & 10 & --\\
14 & 100 & 0 & 100 & 0 & 89 & 9 & 28 & 0 & 37 & 10 & --\\
21 & 97 & 2 & 98 & 2 & 83 & 15 & 28 & 0 & 38 & 12 & 40\\
28 & 86 & 13 & 88 & 11 & 69 & 29 & 26 & 1 & 39 & 11 & 100\\
35 & 63 & 34 & 67 & 30 & 61 & 37 & 26 & 8 & 40 & 9 & 80\\
42 & 52 & 48 & 53 & 47 & 61 & 39 & 32 & 6 & 100 & 0 & 95\\
\bottomrule
\end{tabular}
\caption{Lag density: percentage of the truly compelled arrows oriented correctly and wrongly (generated);
100 series per cell. The remainder are missed or left unoriented.}\label{S-tab:density}
\end{table}

Up to 35 non-zero cells a screen makes orientation no better and mostly worse, because it removes
further lagged links; at 42 cells it gives fewer wrong orientations than the unrestricted search (52 and 39\,\%
against 65 and 48\,\% at $n=500$ and 2000). Reading the
arrows from the contemporaneous graph alone gives few wrong orientations but orients only a quarter to a
third of the arrows. The population graph of the VAR residuals is the same at every density, but the denominator changes: up to
35 non-zero cells the residual search orients about 40\,\% of the 141 to 150 truly compelled arrows correctly
and 9 to 16\,\% wrongly, leaving the rest free; at 42 cells only the 42 arrows identified by the contemporaneous
graph are compelled, and it orients 91 and 100\,\% of them correctly. The unrestricted search has about as few
wrong orientations up to 21 non-zero cells at $n=500$ and 28 at $n=2000$, and more beyond. At 28 non-zero
cells, where the true class compels 148 arrows, it orients 41\,\% wrongly at $n=500$ and 13\,\% at $n=2000$.

\emph{Other lag structures.} Table~\ref{S-tab:designs} keeps the contemporaneous graphs of R7 and changes the lag
structure or the innovation variances: R7 itself; autoregressive terms only at lag 1; R7 with a lag-1
diagonal of 0.7 to 0.85 (dominant root 0.92); R7 with innovation variances log-uniform on $[0.05,5]$; R7's
pattern with coefficients about a third of the size; a full lag-1 matrix; and no dynamics. In the last two
the lags carry no orientation information and 42 of the 150 arrows are compelled. On three designs with a
full lag-1 matrix the fit still omitted one to six lagged links at $n=20\,000$ and $200\,000$ and scored 10 to
33 units better than the true graph. With coefficients a third of the size of R7's, a quarter of the arrows were oriented
wrongly at $n=500$ and a tenth at $n=2000$; unequal variances, strong autoregression and autoregressive terms
only left at most 4\,\% wrong at $n=500$; and with no dynamics, 0.2 to 0.4 spurious lagged links per fit were
enough for 38 to 51\,\% of the unidentified edges to be reported as compelled.

\begin{table}[h]\centering\footnotesize\setlength{\tabcolsep}{4.5pt}
\begin{tabular}{lrrrrrrrr}
\toprule
 & Compelled & \multicolumn{5}{c}{Unrestricted search, extended BIC} & \multicolumn{2}{c}{VAR residuals}\\
Lag structure & in truth & errors & lagged & lagged & right / & free & right / & free\\
 & (of 150) & & missed & spurious & wrong & comp. & wrong & comp.\\
\midrule
\multicolumn{9}{l}{\emph{$n=500$}}\\
R7 as in the paper & 150 & 8.3 & 6.9 & 1.2 & 94 / 5 & -- & 42 / 15 & --\\
own lags only & 150 & 0.5 & 0.0 & 0.4 & 100 / 0 & -- & 43 / 13 & --\\
strong autoregression & 150 & 12.9 & 10.5 & 2.0 & 95 / 4 & -- & 42 / 16 & --\\
unequal innovation variances & 150 & 10.8 & 8.1 & 1.9 & 88 / 4 & -- & 34 / 13 & --\\
weak dynamics & 150 & 30.4 & 28.2 & 1.4 & 69 / 26 & -- & 42 / 13 & --\\
full lag-1 matrix & 42 & 21.4 & 18.4 & 0.8 & 48 / 51 & 91 & 92 / 2 & 39\\
no dynamics & 42 & 1.2 & 0.0 & 0.4 & 94 / 3 & 51 & 93 / 4 & 39\\
\multicolumn{9}{l}{\emph{$n=2000$}}\\
R7 as in the paper & 150 & 0.9 & 0.7 & 0.2 & 100 / 0 & -- & 36 / 8 & --\\
own lags only & 150 & 0.1 & 0.0 & 0.1 & 100 / 0 & -- & 37 / 9 & --\\
strong autoregression & 150 & 3.2 & 2.7 & 0.4 & 100 / 0 & -- & 37 / 10 & --\\
unequal innovation variances & 150 & 3.5 & 2.6 & 0.6 & 97 / 1 & -- & 35 / 11 & --\\
weak dynamics & 150 & 16.7 & 15.6 & 0.9 & 89 / 10 & -- & 38 / 10 & --\\
full lag-1 matrix & 42 & 13.1 & 11.0 & 0.3 & 50 / 50 & 96 & 100 / 0 & 26\\
no dynamics & 42 & 0.6 & 0.0 & 0.2 & 100 / 0 & 38 & 100 / 0 & 29\\
\bottomrule
\end{tabular}
\caption{Other lag structures (generated): mean skeleton errors, lagged links missed and spurious per fit,
percentage of truly compelled arrows oriented correctly and wrongly, and percentage of truly free edges
reported as compelled; 100 series per cell.}\label{S-tab:designs}
\end{table}

%=====================================================================
\section{Unknown lag order}\label{S-sec:order}

The lag order was chosen by AIC, BIC or Hannan--Quinn on the reduced-form VAR (maximum the true order plus
three, common sample) and the unrestricted search with the extended BIC was run with it
(Table~\ref{S-tab:order}). In Model~1 all criteria give the true order in nearly every run. In R7, where the
lag-3 matrix is sparse and weak, the BIC underfits in every run at $n=500$ and the errors nearly triple. At
$k=15$ the BIC chooses one lag in every run. An under-fitted order is far more costly for the graph than a
slightly over-fitted one.

\begin{table}[h]\centering\footnotesize\setlength{\tabcolsep}{4.5pt}
\begin{tabular}{lrrrrrrrll}
\toprule
 & & & True & \multicolumn{4}{c}{Mean errors with the order chosen by} & \multicolumn{2}{c}{Orders chosen
(order:runs)}\\
Design & $n$ & Runs & order & truth & AIC & HQ & BIC & AIC & BIC\\
\midrule
Model 1 & 200 & 200 & 2 & 0.21 & 0.33 & 0.22 & 0.21 & 2:183 3:13 4:3 5:1 & 2:200\\
Model 1 & 500 & 200 & 2 & 0.06 & 0.09 & 0.06 & 0.06 & 2:196 3:4 & 2:200\\
Model 1 & 2000 & 200 & 2 & 0.03 & 0.04 & 0.03 & 0.03 & 2:192 3:6 4:2 & 2:200\\
R7 & 500 & 40 & 3 & 8.62 & 8.78 & 10.68 & 23.50 & 2:1 3:39 & 1:26 2:14\\
R7 & 2000 & 40 & 3 & 1.15 & 1.15 & 1.15 & 2.75 & 3:40 & 1:2 2:2 3:36\\
K15 & 1000 & 20 & 2 & 3.20 & 3.20 & 23.85 & 23.85 & 2:20 & 1:20\\
K15 & 3000 & 20 & 2 & 0.65 & 0.65 & 0.65 & 31.75 & 2:20 & 1:20\\
\bottomrule
\end{tabular}
\caption{Unknown lag order (generated): errors count missing plus extra contemporaneous and lagged
links.}\label{S-tab:order}
\end{table}

%=====================================================================
\section{Application}\label{S-sec:app}

\emph{Bootstrap.} The bootstrap of Figure~1 resamples blocks of eight consecutive rows of the window matrix
$(X_t,X_{t-1},X_{t-2},X_{t-3})$, drawn with replacement and concatenated to the number of window observations,
so that every resampled row is an observed window; the lag order is kept at $p=3$ and the full search with
the extended BIC is rerun on each of 2000 resamples. Resampling blocks of the raw series and forming the
windows afterwards would put a join between unrelated blocks into about three in eight windows. Between 200
and 2000 resamples two of the six arrow frequencies moved by 0.05 to 0.06; at 2000 the standard error of a
frequency near one half is about 0.011. Across block lengths from 4 to 20 the mean adjacency frequency remained between 0.92 and 0.94, while
individual arrow frequencies and their ordering varied (Table~\ref{S-tab:block}): the mean arrow frequency falls
from 0.58 to 0.53 as the block lengthens, the arrows into GDP from 0.74 and 0.77 to 0.61 and 0.67, and
consumption $\to$ GDP, more frequent than unemployment $\to$ bill rate with blocks of 4 (0.74 against 0.59),
is less frequent with blocks of 20 (0.61 against 0.65).

\begin{table}[h]\centering\small
\begin{tabular}{rrrrrrrrr}
\toprule
 & \multicolumn{2}{c}{Mean over six links} & \multicolumn{6}{c}{Frequency of the fitted arrow}\\
Block & adjacency & arrow & C$\to$G & I$\to$G & U$\to$C & U$\to$I & T$\to$P & U$\to$T\\
\midrule
4 & 0.94 & 0.58 & 0.74 & 0.77 & 0.40 & 0.54 & 0.44 & 0.59\\
8 & 0.93 & 0.56 & 0.72 & 0.76 & 0.35 & 0.55 & 0.34 & 0.62\\
12 & 0.92 & 0.53 & 0.67 & 0.72 & 0.31 & 0.57 & 0.33 & 0.63\\
20 & 0.93 & 0.53 & 0.61 & 0.67 & 0.25 & 0.62 & 0.41 & 0.65\\
\bottomrule
\end{tabular}
\caption{Block length: \texttt{gmsvar\_boot()} of version 0.5.0 (blocks of window rows) applied to the fit to
the six US series, 2000 resamples for each length, same seed; the row for blocks of 8 is that of Figure~1. G: GDP, C: consumption, I: investment, P: inflation, T: bill rate, U: unemployment.}\label{S-tab:block}
\end{table}

\emph{Calibration.} To interpret the frequencies, 100 series of $n=202$ were simulated from the fitted
SVAR(3) with Gaussian innovations; each was fitted and bootstrapped in the same way with 200 resamples
(Table~\ref{S-tab:calib}). The fit orients each arrow correctly in 88 to 98\,\% of the series, but the bootstrap
of those series keeps the true arrows in only 64 to 83\,\% of the resamples: bootstrap frequencies measure
stability and understate the reliability of an orientation. Averaged over the six arrows the frequency is
0.72 in the simulations (5th to 95th percentile 0.57 to 0.83) and 0.56 in the data, a value at or below which
2 to 5\,\% of the simulated series fall. The difference lies in the four arrows identified only through lag
exclusions (0.46 in the data against 0.68) and not in the two arrows into GDP (0.74 against 0.82).

\begin{table}[h]\centering\small
\begin{tabular}{llrrr}
\toprule
 & & Fit orients it & \multicolumn{2}{c}{Bootstrap frequency of the arrow}\\
Arrow & identified by & correctly & simulated series & data\\
\midrule
Cons $\to$ GDP & contemporaneous graph & 98\,\% & 0.81 & 0.72\\
Inv $\to$ GDP & contemporaneous graph & 97\,\% & 0.83 & 0.76\\
dUnemp $\to$ dTbill & lag exclusions & 95\,\% & 0.68 & 0.62\\
dUnemp $\to$ Inv & lag exclusions & 88\,\% & 0.67 & 0.55\\
dTbill $\to$ Infl & lag exclusions & 89\,\% & 0.64 & 0.34\\
dUnemp $\to$ Cons & lag exclusions & 95\,\% & 0.71 & 0.35\\
\bottomrule
\end{tabular}
\caption{Calibration of the bootstrap on 100 series simulated from the fitted model.}\label{S-tab:calib}
\end{table}

%=====================================================================


\begin{thebibliography}{99}\small\setlength{\itemsep}{0pt}
\bibitem[Ahelegbey et~al.(2016)]{ahelegbey2016} Ahelegbey, D.~F., Billio, M., and Casarin, R. (2016).
Bayesian graphical models for structural vector autoregressive processes. \emph{Journal of Applied
Econometrics}, 31:357--386.
\bibitem[Andersson et~al.(1997)]{andersson1997} Andersson, S.~A., Madigan, D., and Perlman, M.~D. (1997). A
characterization of Markov equivalence classes for acyclic digraphs. \emph{Annals of Statistics},
25:505--541.
\bibitem[Andrews et~al.(2023)]{andrews2023} Andrews, B., Ramsey, J., Sanchez Romero, R., Camchong, J., and
Kummerfeld, E. (2023). Fast scalable and accurate discovery of DAGs using the best order score search and
grow-shrink trees. In \emph{Advances in Neural Information Processing Systems 36}.
\bibitem[Bang and Didelez(2023)]{bang2023} Bang, C.~W. and Didelez, V. (2023). Do we become wiser with time?
On causal equivalence with tiered background knowledge. In \emph{Proceedings of the 39th Conference on
Uncertainty in Artificial Intelligence}, PMLR 216, pages 119--129.
\bibitem[Basu and Michailidis(2015)]{basu2015} Basu, S. and Michailidis, G. (2015). Regularized estimation in
sparse high-dimensional time series models. \emph{Annals of Statistics}, 43:1535--1567.
\bibitem[Castillo Mansilla and Ninad(2026)]{castillo2026} Castillo Mansilla, I.~G. and Ninad, U. (2026).
Learning causal structure of time series using best order score search. In \emph{Proceedings of the 13th
International Conference on Probabilistic Graphical Models}, PMLR 343, pages 182--209.
\bibitem[Chen and Chen(2008)]{chen2008} Chen, J. and Chen, Z. (2008). Extended Bayesian information criteria
for model selection with large model spaces. \emph{Biometrika}, 95:759--771.
\bibitem[Chernozhukov et~al.(2013)]{chernozhukov2013} Chernozhukov, V., Chetverikov, D., and Kato, K. (2013).
Gaussian approximations and multiplier bootstrap for maxima of sums of high-dimensional random vectors.
\emph{Annals of Statistics}, 41:2786--2819.
\bibitem[Demiralp and Hoover(2003)]{demiralp2003} Demiralp, S. and Hoover, K.~D. (2003). Searching for the
causal structure of a vector autoregression. \emph{Oxford Bulletin of Economics and Statistics}, 65(supplement):745--767.
\bibitem[Drton and Perlman(2007)]{drton2007} Drton, M. and Perlman, M.~D. (2007). Multiple testing and error
control in Gaussian graphical model selection. \emph{Statistical Science}, 22:430--449.
\bibitem[Freedman(1975)]{freedman1975} Freedman, D.~A. (1975). On tail probabilities for martingales.
\emph{Annals of Probability}, 3:100--118.
\bibitem[Hyv\"arinen et~al.(2010)]{hyvarinen2010} Hyv\"arinen, A., Zhang, K., Shimizu, S., and Hoyer, P.~O.
(2010). Estimation of a structural vector autoregression model using non-Gaussianity. \emph{Journal of
Machine Learning Research}, 11:1709--1731.
\bibitem[Kilian and L\"utkepohl(2017)]{kilian2017} Kilian, L. and L\"utkepohl, H. (2017). \emph{Structural
Vector Autoregressive Analysis}. Cambridge University Press.
\bibitem[Larsen et~al.(2025)]{larsen2025} Larsen, T.~E., Ekstr{\o}m, C.~T., and Petersen, A.~H. (2025).
Score-based causal discovery with temporal background information. arXiv:2502.06232.
% CHECK author initials of Ling and Tu before submission
\bibitem[Ling and Tu(2025)]{lingtu2025} Ling and Tu (2025). Variable screening in high-dimensional vector
autoregressions. \emph{Economics Letters}, 257:112695.
\bibitem[L\"utkepohl(2005)]{lutkepohl2005} L\"utkepohl, H. (2005). \emph{New Introduction to Multiple Time
Series Analysis}. Springer.
\bibitem[Malinsky and Spirtes(2018)]{malinsky2018} Malinsky, D. and Spirtes, P. (2018). Causal structure
learning from multivariate time series in settings with unmeasured confounding. In \emph{Proceedings of 2018
ACM SIGKDD Workshop on Causal Discovery}, PMLR 92, pages 23--47.
\bibitem[Meek(1995)]{meek1995} Meek, C. (1995). Causal inference and causal explanation with background
knowledge. In \emph{Proceedings of the Eleventh Conference on Uncertainty in Artificial Intelligence}, pages
403--410.
\bibitem[Moneta(2008)]{moneta2008} Moneta, A. (2008). Graphical causal models and VARs: an empirical
assessment of the real business cycles hypothesis. \emph{Empirical Economics}, 35:275--300.
\bibitem[Nandy et~al.(2018)]{nandy2018} Nandy, P., Hauser, A., and Maathuis, M.~H. (2018). High-dimensional
consistency in score-based and hybrid structure learning. \emph{Annals of Statistics}, 46:3151--3183.
\bibitem[Oxley et~al.(2009)]{oxley2009} Oxley, L., Reale, M., and Tunnicliffe Wilson, G. (2009). Constructing
structural VAR models with conditional independence graphs. \emph{Mathematics and Computers in Simulation},
79:2910--2916.
\bibitem[Pamfil et~al.(2020)]{pamfil2020} Pamfil, R., Sriwattanaworachai, N., Desai, S., Pilgerstorfer, P.,
Georgatzis, K., Beaumont, P., and Aragam, B. (2020). DYNOTEARS: structure learning from time-series data. In
\emph{Proceedings of the 23rd International Conference on Artificial Intelligence and Statistics}, PMLR 108,
pages 1595--1605.
\bibitem[Raskutti and Uhler(2018)]{raskutti2018} Raskutti, G. and Uhler, C. (2018). Learning directed acyclic
graph models based on sparsest permutations. \emph{Stat}, 7:e183.
\bibitem[Reale and Tunnicliffe~Wilson(2001)]{reale2001} Reale, M. and Tunnicliffe Wilson, G. (2001).
Identification of vector AR models with recursive structural errors using conditional independence graphs.
\emph{Statistical Methods and Applications}, 10:49--65.
\bibitem[Runge(2020)]{runge2020} Runge, J. (2020). Discovering contemporaneous and lagged causal relations in
autocorrelated nonlinear time series datasets. In \emph{Proceedings of the 36th Conference on Uncertainty in
Artificial Intelligence}, PMLR 124, pages 1388--1397.
\bibitem[Seabold and Perktold(2010)]{seabold2010} Seabold, S. and Perktold, J. (2010). statsmodels:
econometric and statistical modeling with Python. In \emph{Proceedings of the 9th Python in Science
Conference}, pages 92--96.
\bibitem[Silander and Myllym\"aki(2006)]{silander2006} Silander, T. and Myllym\"aki, P. (2006). A simple
approach for finding the globally optimal Bayesian network structure. In \emph{Proceedings of the 22nd
Conference on Uncertainty in Artificial Intelligence}. AUAI Press.
\bibitem[Slepian(1962)]{slepian1962} Slepian, D. (1962). The one-sided barrier problem for Gaussian noise.
\emph{Bell System Technical Journal}, 41:463--501.
\bibitem[Suter et~al.(2023)]{suter2023} Suter, P., Kuipers, J., Moffa, G., and Beerenwinkel, N. (2023).
Bayesian structure learning and sampling of Bayesian networks with the R package BiDAG. \emph{Journal of
Statistical Software}, 105(9):1--31.
\bibitem[Swanson and Granger(1997)]{swanson1997} Swanson, N.~R. and Granger, C.~W.~J. (1997). Impulse response
functions based on a causal approach to residual orthogonalization in vector autoregressions. \emph{Journal
of the American Statistical Association}, 92:357--367.
\bibitem[Teyssier and Koller(2005)]{teyssier2005} Teyssier, M. and Koller, D. (2005). Ordering-based search:
a simple and effective algorithm for learning Bayesian networks. In \emph{Proceedings of the 21st Conference
on Uncertainty in Artificial Intelligence}. AUAI Press.
\bibitem[van~de Geer and B\"uhlmann(2013)]{vandegeer2013} van~de Geer, S. and B\"uhlmann, P. (2013).
$\ell_0$-penalized maximum likelihood for sparse directed acyclic graphs. \emph{Annals of Statistics},
41:536--567.
\bibitem[Verma and Pearl(1990)]{verma1990} Verma, T. and Pearl, J. (1990). Equivalence and synthesis of causal
models. In \emph{Proceedings of the Sixth Conference on Uncertainty in Artificial Intelligence}, pages
255--270.
\bibitem[Zhang and Wu(2017)]{zhang2017} Zhang, D. and Wu, W.~B. (2017). Gaussian approximation for high
dimensional time series. \emph{Annals of Statistics}, 45:1895--1919.
\end{thebibliography}
\end{document}